\documentclass[article, a4paper, twoside]{memoir}

\counterwithout{section}{chapter}
\setsecnumdepth{subsection}
\setlrmarginsandblock{3.5cm}{2.5cm}{*}
\setulmarginsandblock{2.5cm}{*}{1}
\checkandfixthelayout 

\usepackage[american]{babel}
\usepackage{csquotes}
\usepackage{amsmath,amssymb,amsthm}
\usepackage{bm,bbm}
\usepackage{mathtools}

\usepackage[dvipsnames]{xcolor}
\usepackage{graphicx}
\usepackage{tikz,pgfplots}
\usetikzlibrary{pgfplots.groupplots, arrows}
\tikzstyle{int}=[draw, fill=blue!20, minimum size=2em]
\usepackage{multirow}

\newcommand{\expect}[1]{ \mathbb{E} \left[ #1 \right] }
\newcommand{\cexpect}[2]{ \mathbb{E} \left[ #1 \: \big \vert #2 \: \right] }
\newcommand{\expectund}[2]{ \mathbb{E}_{#2} \left[ #1 \right] }

\newcommand{\proba}[1]{ \mathbb{P} \left( #1 \right) }
\newcommand{\cproba}[2]{ \mathbb{P} \left( #1 \: \big \vert \: #2 \right) }
\newcommand{\cprobaund}[3]{ \mathbb{P}_{#3} \left[ #1 \:\big\vert\: #2 \right] }

\newcommand{\covari}[2]{ \mathbb{C} \text{ov} \left( #1 , #2 \right) }
\newcommand{\ccovari}[3]{ \mathbb{C} \text{ov} \left( #1 , #2 \:\big\vert\: #3 \right) }
\newcommand{\setdef}[1]{ \left\{ #1 \right\} }

\newcommand{\indic}[1]{ \mathbbm{1}_{ \left\{ #1 \right\} } }
\newcommand{\indep}{\rotatebox[origin=c]{90}{$\models$}}
\newcommand{\absval}[1]{ \lvert #1 \rvert }

\usepackage{titling}
\title{Estimating psychometric functions from adaptive designs}
\author{
  Simon Bang Kristensen$^{1,2}$\thanks{
    Correspondance should be sent to Simon Bang Kristensen,
    Odense University Hospital, Heden 16, 5000 Odense, DK-Denmark,
    \href{mailto:sbmkristensen@health.sdu.dk}{sbmkristensen@health.sdu.dk}.
  }
  \and
  Katrine B\o dkergaard$^{1,3}$
  \and
  Bo Martin Bibby$^{1}$ \\[1em]
  $^1$ \small Research Unit for Biostatistics,\\
  \small Department of Public Health, Aarhus University \\
  $^2$ \small Research Unit OPEN, Department of Clinical Research, University of Southern Denmark, \\
  $^3$ \small Department of Clinical Epidemiology, Aarhus University
}

\newcommand{\shortauthor}{Kristensen, B\o dkergaard \& Bibby}
\newcommand{\paperkeywords}{adaptive designs; psychometric function; slope bias}

\date{
}

\usepackage[round]{natbib}
\usepackage{hyperref}
\hypersetup{
  colorlinks = true, linkcolor = black,
  citecolor = black, filecolor = black, urlcolor = blue,
  pdftitle = {\thetitle},
  pdfauthor = {\shortauthor},
  pdfkeywords = {\paperkeywords}
}

\begin{document}

\maketitle

\begin{abstract}
\noindent
An adaptive design adjusts dynamically as information is accrued and a consequence of applying an adaptive design is the potential for inducing small-sample bias in estimates. In psychometrics and psychophysics, a common class of studies investigate a subject's ability to perform a task as a function of the stimulus intensity, meaning the amount or clarity of the information supplied for the task. The relationship between the performance and intensity is represented by a psychometric function. Such experiments routinely apply adaptive designs, which use both previous intensities and performance to assign stimulus intensities, the strategy being to sample intensities where the information about the psychometric function is maximised. Similar schemes are often applied in drug trials to assign doses dynamically using doses and responses from earlier observations. The present paper investigates the influence of adaptation on statistical inference about the psychometric function focusing on estimation, considering both parametric and non-parametric estimation under both fixed and adaptive designs in schemes encompassing within subject independence as well as dependence through random effects. We study the scenarios analytically, focussing on a latent class model to derive results under random effects, and numerically through a simulation study. We show that while the asymptotic properties of estimators are preserved under adaptation, the adaptive nature of the design introduces small-sample bias, in particular in the slope parameter of the psychometric function. We argue that this poses a dilemma for a study applying an adaptive design in the form of a trade-off between more efficient sampling and the need to increase the number of samples to ameliorate small-sample bias.
\end{abstract}

\section{Introduction}
\label{sec:introduction}

A psychometric function is meant to represent a subject's ability to perform a task as a function of the difficulty of the task or the clarity of the presented stimulus. For example, a single trial may ask a participant to identify the geometric shape presented in an image, when the image is shown for a certain number of milliseconds (the intensity of the stimulus). Thus, the stimulus intensity may be thought of as inverse to the task's difficulty. Usually a number of trials are performed over a range of intensities in what may be said to constitute a ``vanilla'' experiment. The purpose of the study is then often centred on some aspect of these experiments, for example comparing performance between groups (e.g. patients and controls), or within person by altering some experimental configuration. As there is a natural upper limit for the time a participant can engage in an experiment a fair amount of attention has focused on making the sampling in the experiment efficient so as to increase the number of configurations, which is usually the focus of the study. In the following we will focus on a ``vanilla'' experiment and the methods to efficiently sample stimulus intensities and we will investigate the consequences to statistical inference, in particular estimation and bias.

The need for flexible sampling leads to the concept of adaptive designs, which is by no means a concept unique to cognitive experiments but to most types of clinical trials. That the design is adaptive means that the design may change dynamically during the trial, usually based on the observed design and possibly outcomes up to a certain time. For example, a \emph{biased coin design} \citep{efron_forcing_1971} tries to make the groups in a randomised trial equally large by adapting the randomisation probability depending on the current allocation of experimental units -- thus using the design but not the outcome at a given step. This is contrary to a classic, fixed design \citep[also see][]{dawid_identifying_2010} where the trial structure is determined before the trial begins -- for example by setting the probability to be allocated to either group to $1/2$ and thereby running the risk of very unbalanced groups when the number of randomised subjects is small. A classic use of outcome adaptive designs occurs in \emph{group sequential designs} \citep[e.g.][]{jennison_group_1999} where one will allow for a trial to stop at an interim stage if the obtained data show strong evidence against the null hypothesis (stopping for efficacy) or if it seems likely that the trial will have a inconclusive outcome (stopping for futility). It is generally accepted that designs using the outcome for example in an interim analysis must account for this in the analysis phase, one problem being the increased type I error rate due to multiple testing, another being the bias of the estimates (confidence intervals must also be produced using special methods). For example, in a drug trial with an interim analysis stopping for efficacy because the observed effect of the drug is very large, the na\"{i}ve estimate of the drug effect will be biased upwards.

In psychometrics and psychophysics, the adaptation uses both the outcome and the design. Simply put, the problem is that if the stimulus intensities are chosen too low or too high, the observed accuracies will be almost all zeros or ones leading to poor estimates of the psychometric function. Thus, we would like to sample where there is more information about the psychometric function. As these accuracies are specific to a participant and there is no way to know the participant's level of accuracy before the experiment, the design needs to adapt based on the previous intensities along with the previous performance. Various designs have been proposed as reviewed for example in \cite{treutwein_adaptive_1995} and \cite{leek_adaptive_2001}, the simplest of which adapt the stimulus intensity at a given time from the previous intensity based on the performance a few trials back.

Estimates from adaptive designs will usually inherit the asymptotic properties of those from the fixed sample design \citep[e.g.][]{melfi_estimation_2000}. However, this does not account for the behaviours in small samples and while there is acknowledgement for the need to account for these in the medical trial literature this does not seem common in psychometrics (cf. for instance section 3.4.2 and 5.4.2 of the textbook \cite{kingdom_psychophysics:_2016}). As also highlighted by \cite{bretz2009adaptive} testing and the control of type I error rates are much better understood than estimation in adaptive designs.

The article proceeds as follows. We first give a brief overview of some standard methods for adaptive stimulus allocation. We then introduce the basic setup for the paper introducing the concepts of psychometric functions along with notation for designs and dependence schemes, before investigating the likelihoods as the basis of inference and estimation. We perform these investigations both in designs employing adaptive and non-adaptive allocation as well as under within-subject independence and dependence through the inclusion of random effects. Finally, we illustrate these points through a simulation study followed by a brief discussion.

\subsection{Brief overview of adaptive methods}
\label{sec:overv-adapt-meth}

In the following we give a brief review of methods used in psychometrics and psychophysics for constructing adaptive designs \citep[see][for a more complete and in-depth treatment]{treutwein_adaptive_1995, leek_adaptive_2001}. As there is substantial overlap with the literature on dose finding designs, we include a few references from the related literature and make some comparisons.

The simplest class of adaptive designs assigns the next stimulus intensity from the current intensity based on the performance of the subject a few trials back. This includes the up-down design \citep{dixon_method_1948} in which the stimulus is increased from the current intensity if the current response was incorrect and decreased if the response was correct. This procedure targets the accuracy probability $1/2$. If the purpose of the study is to estimate some other quantile of the psychometric function, this may constitute a disadvantage. Other designs may be employed to target different accuracies, e.g. the one-up-two-down design will decrease the intensity only after two consecutive correct responses and target the probability $1/\sqrt{2} = 0.71$.
A more general approach are the so-called weighted up-down methods, where rather than altering the number of correct responses before decreasing the intensity, the weighted up-down designs employ differential up and down stepsizes depending on the target probability. Note that these designs are all characterised by the fact that they will find the intensity corresponding to the target probability and then fluctuate around this level, not converging. Moreover, the stepsize when adjusting the intensity is fixed across trials.

Stochastic approximation schemes similarly use differential up and down step sizes depending on the target probability but further decreases the step sizes with the number of trials. This leads to convergence to the target probability. 

A more involved class of adaptive methods utilises a broader range of the information from the previous trials. The method may be summarised as performing maximum likelihood estimation of the parameters of the psychometric function following each trial using all available information on responses and stimulus intensities up to that point. Based on predictions from these estimates, a next stimulus intensity is chosen that in some way adds the most information. The procedure is known in psychophysics as ``best PEST'' following \cite{pentland_maximum_1980}, who proposed it building on a series of more ad hoc methods known as PEST (Parameter Estimation by Sequential Testing). 

In the dose-finding literature, the adaptive methods also commonly include up-down type designs as well as stochastic approximation and sequential maximum likelihood is also often done, see \cite{oquigley_methods_1991} for a review. There, the latter method is termed the continual reassessment method (CRM) \citep{oquigley_continual_1990, oquigley_continual_1996}.   

In proposing the CRM, \cite{oquigley_continual_1990} suggest the use of a prior (advocating for a weak prior) on the does-response parameters, thus setting the method in a Bayesian framework so that the next stimulus intensity is determined by maximising the posterior distribution rather than the likelihood. The ``best PEST'' does not make such assumptions, but requires a burnin period or reverting to other methods when likelihood estimation fails. This is realistic in most psychophysics procedures where there are typically many trials per subject, but not so in designs such as those considered by \cite{oquigley_continual_1990}, where there may only be 20 patients included, each tested once. Thus, the adding of a prior lends additional numerical stability.

A Bayesian version of the ``best PEST'' is given by \cite{watson_quest:_1983}. \cite{kontsevich_bayesian_1999} further study a problem of estimating the threshold while regarding the slope parameter as a ``nuisance'' parameter. Targeting specific parameters of the psychometric function while accounting for the other parameters is related to so-called psi-methods, which we do not elaborate on here \citep[see][Section 5.4 for an overview]{kingdom_psychophysics:_2016}.

\section{Basic theory}
\label{sec:basic-theory}

We first introduce some notation. Let an observation be of the form $(Y, S)$ where $Y \in \setdef{0,1}$ is the accuracy and $S$ is the stimulus intensity taking values in a finite set $\mathcal{S}$. Suppose that intensities are equidistant and we will assume without loss of generality that $\mathcal{S}=\setdef{1,\ldots, D}$. Observations are taken for $i=1,\ldots, N$ subjects at $t=1,\ldots,T$ time points leading to the data $\setdef{(Y_{it}, S_{it})}$. Note that we simplistically assume that $T$ is fixed, meaning that there is no data dependent stopping, and that all subjects participate in the same number of trials. Write $(\bm{Y}_i, \bm{S}_i)$ for the combined vector of observations on subject $i$. Further, when $W$ is some random variable we will write $f_W$ for its density.

\subsection{Psychometric functions}
\label{sec:psych-funct}

It is convenient first to regard the design as being fixed so that we may either consider the intensities as fixed or as ancillaries (see below). A simple psychometric function model for the accuracy given the intensity would be,
\begin{equation}
  \label{eq:1}
  M_{(a)}: \quad Y_{it} \: \big\vert \: S_{it}
  \sim \text{b} \left(1, F(S_{it} ; \theta) \right) ,
\end{equation}
so that the accuracy is a Bernoulli variable with a success probability depending on the stimulus intensity. We assume that subjects are independent but postpone specification of the within-subject dependence. The function $F$ is the psychometric function, which establishes the relationship between the success probability and the intensity. $\theta$ denotes the parameters of the conditional distribution of the response given the intensity, which is assumed to be the parameters of interest. Let $\Omega_{\theta}$ be the domain for $\theta$. In a simple case, $F$ could be logistic $x \mapsto \left(1 + e^{(x - \tilde{a})/\tilde{b}}\right)^{-1} = \left(1 + e^{-\left(a + b x\right)}\right)^{-1}$, in which case $\theta = (a,b)$ would be the intercept and slope on the logistic scale. Further, we will consider a random effects model,
\begin{equation}
  \label{eq:51}
  M_{(b)}: \quad Y_{it} \: \big\vert \: S_{it}, \bm{\alpha}_i
  \sim \text{b} \left(1, F(S_{it} ; \bm{\alpha}_i, \theta) \right) ,
\end{equation}
where $\bm{\alpha}_i$ is the random effects for subject $i$. To fix ideas we will consider a logistic model with random intercepts,
\begin{equation}
  \label{eq:6}
  F(s;\theta, \alpha_i) = \frac{1}{1 + e^{- \left(a + \alpha_i + b \cdot s \right)}}, \quad
  \alpha_i \sim N(0, \tau^2) ,
\end{equation}
where $\theta = (a,b)$.

\subsubsection{Parametric and non-parametric models}
\label{sec:param-non-param}

We will in the following discern two scenarios \citep[as also done in][]{treutwein_adaptive_1995}: The \emph{parametric} scenario where, as outlined above, we assume a specific form of the psychometric function with the objective of estimating the parameters, for example the logistic function in (\ref{eq:6}). We will also consider the \emph{non-parametric} scenario, where we wish to estimate accuracy probabilities at given intensities, i.e. $\pi_s = F(s)$ for $s \in \mathcal{S}$ and some unknown function $F$.

\subsection{Fixed designs}
\label{sec:fixed-designs}

The simplest approach to choosing the stimulus intensity in a series of trials is to take them to be random in the set $\mathcal{S}$ with distribution $f_S$. For example, the intensities may be sampled to be uniformly random or could be Gaussian to sample the majority of intensities close to some prespecified intensity. As the intensities in this case may simply be determined before the trial, we term this the \emph{fixed design}, as the intensities are determined before the study begins. We will denote by $\psi \in \Omega_{\psi}$ any parameters of the distribution of the stimulus intensities.  

\subsection{The up-down design}
\label{sec:up-down-design}

As described in Section \ref{sec:overv-adapt-meth} the up-down design samples the stimulus in a given trial based on the intensity and performance of the previous trial. Recall that the intensities take values in $\mathcal{S}=\setdef{1,\ldots, D}$. For participant $i$, randomise $S_{i1}$ uniformly in $\mathcal{S}$ and set
\begin{equation}
  \label{eq:3}
  S_{it} =
  \begin{cases}
    1 & \text{if } S_{i (t-1)} - \left[ 2 \cdot Y_{i (t-1)} - 1 \right]  = 0 \\
    S_{i (t-1)} - \left[ 2 \cdot Y_{i (t-1)} - 1 \right] & \text{else} \\
    D & \text{if } S_{i (t-1)} - \left[ 2 \cdot Y_{i (t-1)} - 1 \right] = D+1
  \end{cases}
\end{equation}
for $t=2, \ldots, T$. The resulting allocation of the stimulus intensities is called the \emph{up-down design}.

\subsection{Individual versus group level analyses}
\label{sec:indiv-vers-group}

In the following we limit our discussion to an analysis strategy in which the aim is to analyse the entire set of data once it is collected, which we refer to as a \emph{group level} analysis. In practice, another strategy is sometimes applied in which the data is analysed on an individual level to obtain estimates $\hat{\theta}_1, \ldots, \hat{\theta}_N$ which are then analysed, usually in a multiple regression model. In the vanilla experiment described in the Introduction where we only consider data collected from one experimental configuration under various intensities, we might in the second stage be interested in the expectation of the parameter of interest, and take the estimator simply to be the mean (analogously to an intercept-only ``multiple'' regression),
\begin{equation}
  \label{eq:54}
  \hat{\theta} = \frac{1}{N} \sum_{i=1}^{N} \hat{\theta}_i .
\end{equation}
Naturally, this two-stage estimator is less efficient than the one that arises from using all the data. Additionally, it tacitly relies on an assumption of normality of the estimates around the true parameter, which may not be true in small samples. Indeed, as we shall elaborate on, if an adaptive design is applied, the adaptation will imply a bias in small samples even in cases that one would expect to be unbiased as in the fixed design. The distinction between group and individual level analysis bears importance as the random effects model introduced in Section \ref{sec:psych-funct} models between-participant variation while enabling estimation of the parameters of interest, but the random effects would not be identifiable in the individual level analysis. 

Finally, note that the individual level estimates are used by the continual reassessment method described in Section \ref{sec:overv-adapt-meth} to update the stimulus intensity. We would still, however, apply an analysis of all the data once it has been collected in its entirety.

\subsection{Dependence schemes}
\label{sec:two-depend-schem}

Below we introduce within-subject dependence schemes, which depend on the basic model for the accuracy given the stimulus intensity as introduced in Section \ref{sec:psych-funct} as well as the design of the stimulus allocation. We consider the usual fixed design case along with the up-down design with no random effects as well as the fixed and up-down designs with random effects. We will use so-called directed acyclic graphs (DAGs) to represent the dependence structure in the various schemes. The theory of such graphs is well developed \citep[e.g.][]{lauritzen_graphical_1996} and we do not presume to reiterate it here. Rather, we will rely on intuitive arguments based on these. A more formal algorithm for reading conditional independence from a DAG is given in Appendix \ref{cha:cond-indep-from}. An important assumption is that the graph accurately encodes the dependence of the scheme.

\subsubsection{Scheme 1: Fixed design ($\text{FD}$ and $\text{FD}_r$)}
\label{sec:scheme-1}

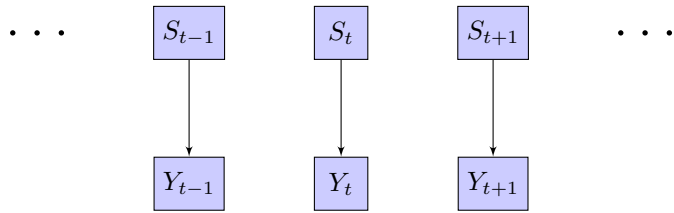
\begin{figure}[htb]
  \centering
  \begin{tikzpicture}[node distance=2.5cm,auto,>=latex']
    \node [int] (s1) {$S_{t-1}$};
    \node [int, below of = s1, node distance=2cm] (y1) {$Y_{t-1}$};
    \node [int, right of = s1, node distance=2cm] (s2) {$S_t$};
    \node [int, below of = s2, node distance=2cm] (y2) {$Y_t$};
    \node [int, right of = s2, node distance=2cm] (s3) {$S_{t+1}$};
    \node [int, below of = s3, node distance=2cm] (y3) {$Y_{t+1}$};
    \node [right of = s3, node distance=2cm, minimum size=2em] (empt) {\Huge $\ldots$};
    \node [left of = s1, node distance=2cm, minimum size=2em] (empt) {\Huge $\ldots$};
    \path[->] (s1) edge node {} (y1);
    \path[->] (s2) edge node {} (y2);
    \path[->] (s3) edge node {} (y3);
  \end{tikzpicture}
  \caption{Scheme FD: Fixed design with no random effects.}
  \label{fig:scheme-1a}
\end{figure}

We first consider a model, where a single observation follows the simple psychometric function model $M_{(a)}$ in (\ref{eq:1}). The intensities are sampled using a design that was fixed at the beginning of the experiment with no adaptation, see Section \ref{sec:fixed-designs}. Denote this by $\text{FD}$, as illustrated in Figure \ref{fig:scheme-1a}.
 
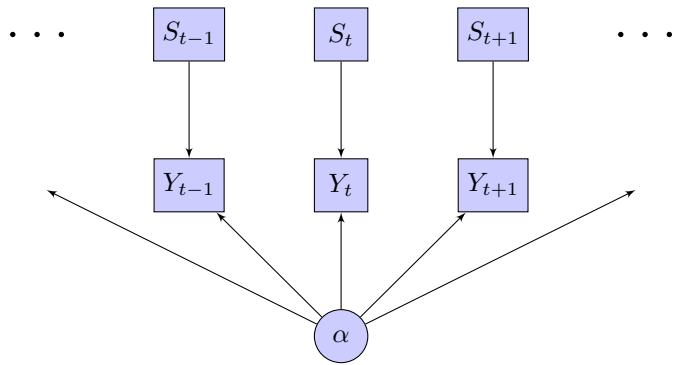
\begin{figure}[htb]
  \centering
  \begin{tikzpicture}[node distance=2.5cm,auto,>=latex']
    \node [int] (s1) {$S_{t-1}$};
    \node [int, below of = s1, node distance=2cm] (y1) {$Y_{t-1}$};
    \node [int, right of = s1, node distance=2cm] (s2) {$S_t$};
    \node [int, below of = s2, node distance=2cm] (y2) {$Y_t$};
    \node [int, right of = s2, node distance=2cm] (s3) {$S_{t+1}$};
    \node [int, below of = s3, node distance=2cm] (y3) {$Y_{t+1}$};
    \node [right of = s3, node distance=2cm, minimum size=2em] (empt) {\Huge $\ldots$};
    \node [left of = s1, node distance=2cm, minimum size=2em] (empt2) {\Huge $\ldots$};
    \node [below of = empt, node distance=2cm] (empty) {};
    \node [below of = empt2, node distance=2cm] (empty2) {};
    \node [int, circle, below of = y2, node distance=2cm] (alph) {$\alpha$};
    \path[->] (s1) edge node {} (y1);
    \path[->] (s2) edge node {} (y2);
    \path[->] (s3) edge node {} (y3);
    \path[->] (alph) edge node {} (y1);
    \path[->] (alph) edge node {} (y2);
    \path[->] (alph) edge node {} (y3);
    \path[->] (alph) edge node {} (empty);
    \path[->] (alph) edge node {} (empty2);
  \end{tikzpicture}
  \caption{Scheme $\text{FD}_r$: Fixed design with random effects.}
  \label{fig:scheme-1b}
\end{figure}

We also use the same scheme but under the assumption that the observations are conditionally independent given the intensity and a subject level random effect, cf. the model $M_{(b)}$ in (\ref{eq:51}). We call this $\text{FD}_r$, where the subscript indicates the presence of a random effect, see Figure \ref{fig:scheme-1b}.

\subsubsection{Scheme 2: Up-down design ($\text{UD}$ and $\text{UD}_r$)}
\label{sec:scheme-2}

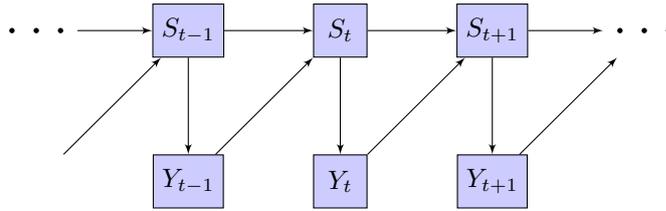
\begin{figure}[htb]
  \centering
  \begin{tikzpicture}[node distance=2.5cm,auto,>=latex']
    \node [int] (s1) {$S_{t-1}$};
    \node [int, below of = s1, node distance=2cm] (y1) {$Y_{t-1}$};
    \node [int, right of = s1, node distance=2cm] (s2) {$S_t$};
    \node [int, below of = s2, node distance=2cm] (y2) {$Y_t$};
    \node [int, right of = s2, node distance=2cm] (s3) {$S_{t+1}$};
    \node [int, below of = s3, node distance=2cm] (y3) {$Y_{t+1}$};
    \node [right of = s3, node distance=2cm, minimum size=2em] (empt) {\Huge $\ldots$};
    \node [left of = s1, node distance=2cm, minimum size=2em] (empt2) {\Huge $\ldots$};
    \node [below of = empt2, node distance=2cm, minimum size=2em] (empty2) {};
    \path[->] (s1) edge node {} (y1);
    \path[->] (s1) edge node {} (s2);
    \path[->] (y1) edge node {} (s2);
    \path[->] (s2) edge node {} (y2);
    \path[->] (s2) edge node {} (s3);
    \path[->] (y2) edge node {} (s3);
    \path[->] (s3) edge node {} (y3);
    \path[->] (s3) edge node {} (empt);
    \path[->] (y3) edge node {} (empt);
    \path[->] (empt2) edge node {} (s1);
    \path[->] (empty2) edge node {} (s1);
  \end{tikzpicture}
  \caption{Scheme UD: Up-down design.}
  \label{fig:scheme2a}
\end{figure}

Consider also the basic model $M_{a}$ in (\ref{eq:1}) but with an up-down design as described in Section \ref{sec:up-down-design}. This is the scheme $\text{UD}$, as illustrated in Figure \ref{fig:scheme2a}. Here, the stimulus intensity is determined at time $t$ by the previous intensity along with the previous response.

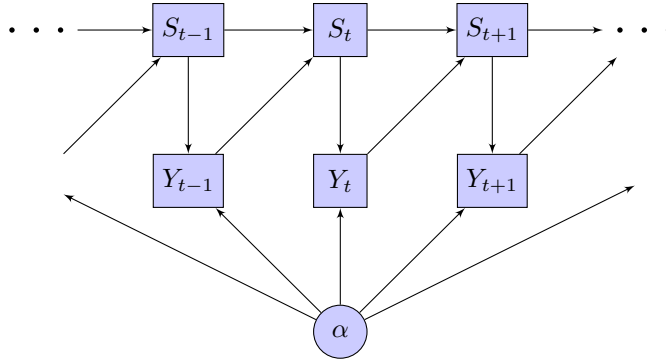
\begin{figure}[htb]
  \centering
  \begin{tikzpicture}[node distance=2.5cm,auto,>=latex']
    \node [int] (s1) {$S_{t-1}$};
    \node [int, below of = s1, node distance=2cm] (y1) {$Y_{t-1}$};
    \node [int, right of = s1, node distance=2cm] (s2) {$S_t$};
    \node [int, below of = s2, node distance=2cm] (y2) {$Y_t$};
    \node [int, right of = s2, node distance=2cm] (s3) {$S_{t+1}$};
    \node [int, below of = s3, node distance=2cm] (y3) {$Y_{t+1}$};
    \node [right of = s3, node distance=2cm, minimum size=2em] (empt) {\Huge $\ldots$};
    \node [below of = empt, node distance=2cm] (empty) {};
    \node [left of = s1, node distance=2cm, minimum size=2em] (empt2) {\Huge $\ldots$};
    \node [below of = empt2, node distance=2cm, minimum size=2em] (empty2) {};
    \node [int, circle, below of = y2, node distance=2cm] (alph) {$\alpha$};
    \path[->] (s1) edge node {} (y1);
    \path[->] (s1) edge node {} (s2);
    \path[->] (y1) edge node {} (s2);
    \path[->] (s2) edge node {} (y2);
    \path[->] (y2) edge node {} (s3);
    \path[->] (s2) edge node {} (s3);
    \path[->] (s3) edge node {} (y3);
    \path[->] (s3) edge node {} (empt);
    \path[->] (y3) edge node {} (empt);
    \path[->] (alph) edge node {} (y1);
    \path[->] (alph) edge node {} (y2);
    \path[->] (alph) edge node {} (y3);
    \path[->] (alph) edge node {} (empty);
    \path[->] (alph) edge node {} (empty2);
    \path[->] (empt2) edge node {} (s1);
    \path[->] (empty2) edge node {} (s1);
  \end{tikzpicture}
  \caption{Scheme $\text{UD}_r$: Up-down design with random effect $\alpha$.}
  \label{fig:scheme2b}
\end{figure}

As above we additionally consider the same scheme including a random effect, cf. Figure \ref{fig:scheme2b} illustrating this scheme, which we denote $\text{UD}_r$.

\section{Inference for fixed and adaptive design}
\label{sec:infer-fixed-adapt}

Recall that $\theta$ contains the parameters of interest and all the parameters of the conditional distribution of the response $Y$ given the intensities $S$. In the fixed design, the distribution of the stimulus intensities $f_S$ depends only on some auxiliary parameter $\psi$, for example if the intensity is determined from randomisation using these parameters, and we will assume that this is functionally independent from the parameter of interest so that $(\theta, \psi) \in \Omega_{(\theta, \psi)} = \Omega_{\theta} \times \Omega_{\psi}$. This means that $S$ is ancillary for $\theta$ \citep[e.g.][]{barndorff-nielsen_inference_1994}.

We will perform estimation by maximising the likelihood function. Since the subjects are independent we focus on the likelihood contribution from a single subject. Under the fixed design, the likelihood contribution from subject $i$ is given by,
\begin{equation}
  \label{eq:2}
  \begin{aligned}
  L_i (\theta, \psi) &= f_{(\bm{Y}_i, \bm{S}_i)} (\bm{y}_i, \bm{s}_i; \theta,\psi) \\
  &= f_{\bm{Y}_i \: \vert \: \bm{S}_i} (\bm{y}_i \: \big\vert \: \bm{s}_i; \theta)
  f_{ \bm{S}_i} (\bm{s}_i; \psi) .
\end{aligned}
\end{equation}
Note that the estimate arising from this likelihood (by differentiation of the log-likelihood) is the same as that using $L_i (\theta) = f_{\bm{Y}_i \: \vert \: \bm{S}_i}$. This is the usual argument for conditionality -- inference should take place conditional on the ancillary statistic $S$ since this will not affect inference about $\theta$, but we get rid of the nuisance parameter $\psi$. This is why, in the fixed design, we could also simply think of the stimulus intensities as being non-random, fixed quantities.

In an adaptive design, this no longer holds. Since $S$ is chosen based on the performance it will not be ancillary for $\theta$, but carry information about the subject's psychometric function. Consider the simple up-down design introduced above: If we momentarily ignore the boundaries on the intensity (i.e. take $\mathcal{S} = \setdef{\ldots, -2,-1,0,1,2,\ldots} = \mathbb{Z}$ to be the integers) we may derive the following recursive identity for the probability mass function of the intensity,
\begin{equation}
  \label{eq:4}
  f_{S_{it}} (s) = F(s+1;\theta) f_{S_{i (t-1)}} (s+1) +
  \left[ 1- F(s-1;\theta) \right] f_{S_{i (t-1)}} (s-1) ,
  \quad s \in \mathcal{S} ,
\end{equation}
for $t = 2, \ldots, T$. Note that (\ref{eq:4}) expresses that there are two ways to observe the intensity $s$ at trial $t$ -- the intensity was $s+1$ in the previous trial and a correct response was given (which happens with probability $F(s+1;\theta)$) so that the intensity is decreased, or the previous intensity was $s-1$ and an incorrect response was given (with probability $1-F(s-1;\theta)$) leading to the intensity being increased. In short, (\ref{eq:4}) shows that the distribution of $S_{it}$ (for $t \geq 2$) will depend on $\theta$.

As we will show below, however, the likelihood function is unchanged under most realistic adaptive designs, showing that the asymptotic properties of the estimators are also unchanged. However, bias may be introduced in small samples. A similar problem is known in the theory of group sequential designs. Here it is possible to find sufficient statistics in most models (this is normally the ``usual'' sufficient statistic, e.g. the sum, along with the stopping time), but due to the non-random stoppage the statistic is not complete, and thus the usual invocation of the Lehmann-Scheff\'{e} theorem does not produce a best unbiased estimator \citep{milanzi_estimation_2015}. This can be tied to the problem of non-ancillarity \citep{kenward_likelihood_1998}. In group sequential trials, several adjusted estimators have been proposed to remedy the bias. \cite{whitehead_bias_1986} proposed a bias corrected estimator, while other authors \citep[e.g.][]{fan_conditional_2004, milanzi_estimation_2015} consider estimators based on the conditional likelihood given the stopping time. \cite{fan_conditional_2004} argue that it is more reasonable to focus on the conditional bias given the stopping time, as the bias is usually large and positive (for positive treatment effect) for early stopping, while it is slightly negatively biased for later stopping leading to a, by comparison, small overall bias. They show that the maximum conditional likelihood estimator is equivalent to a conditional bias adjusted estimator. It is less clear, however, how this approach should be applied in an adaptive design, whether one should condition on the intensity at a concurrent time or the design in its entirety. I.e. it is not obvious if the basic likelihood quantity in the conditional approach should be $f_{Y_{it} \:\vert\: S_{it}}$ or $f_{\bm{Y}_{i} \:\vert\: \bm{S}_{i}}$. Another estimator proposed in the group sequential design only uses the data up until the first interim, which is unbiased as it is independent of any look at the data. The accompanying high variance is ameliorated for example by applying Rao-Blackwellisation conditioning on the sufficient statistics mentioned above. This technique does not easily extend to adaptive designs as the adaptation begins after the first trial thus supplying no information on the psychometric function independent of the data look. \cite{bowden_unbiased_2017} propose an inverse probability weighted estimator for an adaptive design to which they also apply Rao-Blackwellisation.

\subsection{Estimation under $\text{FD}$ and $\text{UD}$ scheme}
\label{sec:estim-under-textfd}

When assuming conditional within-participant independence given the previous intensities our setup is the same as that studied in \cite{bowden_unbiased_2017}. They consider a case where independent individuals are randomized into a trial, so the independence assumption is well warranted there.

Adapting their arguments to our present setup we may decompose an individual likelihood contribution (dropping the $i$ subscript for convenience), writing $\bm{Y}_{-t} = (Y_1, \ldots, Y_{t-1})$,
\begin{equation}
  \label{eq:7}
  \begin{aligned}
    L (\theta, \psi) &= \prod_{t=1}^{T} f_{(Y_t,S_t) \: \vert \: \bm{Y}_{-t},\bm{S}_{-t}} \\
    &=
    \prod_{t=1}^{T} f_{Y_t \: \vert \: \bm{Y}_{-t},\bm{S}_{-t}, S_t} \cdot
    f_{S_t \: \vert \: \bm{Y}_{-t},\bm{S}_{-t}} .
  \end{aligned}
\end{equation}
The second factor in the product may reasonably be assumed not to depend on $\theta$ as it is the distribution of the intensity given all previous information on both intensities and outcomes. Thus, as before we may drop this factor and base our estimation instead on the likelihood arising from the contributions,
\begin{equation}
  \label{eq:8}
  L (\theta) = \prod_{t=1}^{T} f_{Y_t \: \vert \: \bm{Y}_{-t},\bm{S}_{-t}, S_t}
  (y_t \: \vert \: \bm{y}_{-t},\bm{s}_{-t}, s_t) .
\end{equation}
Under scheme FD and UD we have conditional independence of the current outcome from earlier outcomes as well as intensities given the current intensity. To see this for scheme UD, we note from the DAG in Figure \ref{fig:scheme2a} that $Y_t \: \vert \: \bm{Y}_{-t},\bm{S}_{-t}, S_t \sim Y_t \: \vert \: S_t$, for example using the algorithm detailed in Appendix \ref{cha:cond-indep-from}. Using this and inserting the psychometric function from $M_{(a)}$ in (\ref{eq:1}),
\begin{equation}
  \label{eq:9}
  \begin{aligned}
    L (\theta) &=
    \prod_{t=1}^{T} f_{Y_t \: \vert \: S_t} (y_t \: \vert \: s_t) \\
    &=
    \prod_{t=1}^{T} F(s_t;\theta)^{y_t} \left[1 - F(s_t;\theta) \right]^{1 - y_t} ,
  \end{aligned}
\end{equation}
and then the maximum likelihood estimate will coincide with that given under the fixed design.

A crucial observation is the difference to the usual estimate in terms of small sample performance owing to stochastic sampling. This will be more apparent if we consider estimating the accuracy probabilities themselves rather that $\theta$. To this end, imagine a scenario where we were unwilling to assume a specific form of $F$ and instead would simply estimate $\pi_s = F(s)$ for $s \in \mathcal{S}$, i.e. the non-parametric setup. Then the log-likelihood from (\ref{eq:9}) for all $N$ subjects is, 
\begin{equation}
  \label{eq:10}
  \begin{aligned}
    l (\theta) &=
    \sum_{i=1}^{N}
    \sum_{t=1}^{T}
    \left[
      y_{it} \log F(s_{it};\theta) + (1 - y_{it}) \log \left( 1 - F(s_{it};\theta) \right)
    \right] \\
    &=
    \sum_{s \in \mathcal{S}}
    \left[
      m_s \log \pi_s + (T_s - m_s) \log \left( 1 - \pi_s \right)
    \right] ,
\end{aligned}
\end{equation}
where $m_s$ is the number of accurate answers under intensity $s$ and $T_s$ is the total number of trials under intensity $s$. Note that $T_s$ is a random variable. Differentiating (\ref{eq:10}) we obtain the ``usual'' estimates,
\begin{equation}
  \label{eq:11}
  \widehat{\pi}_s = \frac{m_s}{T_s}
  = \frac{
    \sum_{i=1}^{N} \sum_{t: S_{it} = s} Y_{it} 
  }{
    \sum_{i=1}^{N} \sum_{t=1}^{T} \indic{S_{it} = s}  
  }
\end{equation}
Note that if the denominator in (\ref{eq:11}) were fixed we would have $\expect{\widehat{\pi}_s} = \cproba{Y_{it} = 1}{S_{it} = s} = \pi_s$ meaning that the estimator is unbiased. However, as already noted, $T_s$ is stochastic. Adapting the argument from Section 3 in \cite{bowden_unbiased_2017}, we have that
\begin{equation}
  \label{eq:16}
  \begin{aligned}
    \expect{T_s \hat{\pi}_s} - \pi_s \expect{T_s} &=
    \expect{\left(\hat{\pi}_s - \pi_s \right) T_s} \\
    &=
    \expect{m_s - \pi_s T_s} \\
    &=
    \sum_{i=1}^{N} \sum_{t=1}^{T} \left[
      \proba{S_{it} = s, Y_{it} = 1}
      - \cproba{Y_{it} = 1}{S_{it} = s}  \proba{S_{it} = s}
    \right] \\
    &= 0 ,
  \end{aligned}
\end{equation}
and adding and subtracting $\expect{T_s} \expect{\hat{\pi}_s}$ it follows that
\begin{equation}
  \label{eq:12}
  \covari{T_s}{\hat{\pi}_s} + \expect{T_s} \expect{\hat{\pi}_s} - \pi_s \expect{T_s}
  = 0 ,
\end{equation}
which in turn implies,
\begin{equation}
  \label{eq:13}
  \text{Bias} (\hat{\pi}_s) = - \frac{\covari{T_s}{\hat{\pi}_s}}{\expect{T_s}} .
\end{equation}

The bias in formula (\ref{eq:13}) concurs with the intuition that the bias in the estimator depends on the degree to which the estimator influences the design. This may also correspond to the following intuition. Imagine a study employing an adaptive design to target the centre of an underlying psychometric function. Suppose that the success probability $\pi_s$ at intensity $s$ is fairly high but is also over-estimated in the beginning of the study due to random variation. We would then expect to be close to the upper edge of the psychometric function and the adaptive design would compensate by favouring more samples at lower intensities. Consequently we may not get enough additional samples at intensity $s$ to rectify our overestimate of $\pi_s$. In this case, the covariance between $T_s$ and $\hat{\pi}_s$ would by negative (a higher estimate of the success probability at $s$ would lead to a lower number of total trials at $s$) and formula (\ref{eq:13}) implies that the bias is positive, as expected. Conversely, if we underestimate a low success probability, the covariance between the estimates and the total would be positive and (\ref{eq:13}) confirms the downwards bias. We also remark that the bias is decreased as more observations are expected at stimulus intensity $s$.

\subsection{Estimation under schemes $\text{FD}_r$ and $\text{UD}_r$}
\label{sec:estim-under-textfd_r}

If we apply the same likelihood decomposition as above in (\ref{eq:7}), we may write the likelihood contribution for subject $i$ under $\text{FD}_r$ and $\text{UD}_r$ as,
\begin{equation}
  \label{eq:52}
  \begin{aligned}
    L_i(\theta, \tau, \psi)
    &=
    \int_{}^{}
    f_{\alpha_i} \cdot
    f_{(\bm{Y}_i , \bm{S}_i) \: \vert \: \alpha_i}
    \: d \alpha_i \\
    &=
    \int_{}^{}
    f_{\alpha_i} \cdot
    \prod_{t=1}^{T} f_{(Y_{it},S_{it}) \: \vert \: \bm{Y}_{i -t},\bm{S}_{i -t}, \alpha_i}
    \: d \alpha_i \\
    &=
    \int_{}^{}
    f_{\alpha_i} \cdot
    \prod_{t=1}^{T} f_{Y_{it} \: \vert \: \bm{Y}_{i -t},\bm{S}_{i -t}, S_{it}, \alpha_i}
    \cdot
    f_{S_{it} \: \vert \: \bm{Y}_{i -t},\bm{S}_{i -t}, \alpha_i}
    \: d \alpha_i .
  \end{aligned}
\end{equation}
The key is then to note that if the dependence structure is indeed accurately represented by the DAG in Figure \ref{fig:scheme-1b} (scheme $\text{FD}_r$) or \ref{fig:scheme2b} (scheme $\text{UD}_r$), then $S_{it} \indep \alpha_i \: \vert \: \bm{Y}_{i -t},\bm{S}_{i -t}$. Thus, the likelihood may be reduced since the conditional distribution of $S_{it}$ given $\bm{Y}_{i -t}$ and $\bm{S}_{i -t}$ does not depend on the parameter of interest, so that,
\begin{equation}
  \label{eq:53}
  \begin{aligned}
    L_i(\theta, \tau, \psi) &=
    \prod_{t=1}^{T} f_{S_{it} \: \vert \: \bm{Y}_{i -t},\bm{S}_{i -t}}
    \int_{}^{}
    f_{\alpha_i} \cdot
    \prod_{t=1}^{T} f_{Y_{it} \: \vert \: \bm{Y}_{i -t}, \bm{S}_{i -t}, S_{it}, \alpha_i}
    \: d \alpha_i \\
    &\propto
    \int_{}^{}
    f_{\alpha_i} \cdot
    \prod_{t=1}^{T} f_{Y_{it} \: \vert \: \bm{Y}_{i -t} , \bm{S}_{i -t}, S_{it}, \alpha_i}
    \: d \alpha_i \\
    &=
    \int_{}^{}
    f_{\alpha_i} \cdot
    \prod_{t=1}^{T} f_{Y_{it} \: \vert \: S_{it}, \alpha_i}
    \: d \alpha_i \\
    &=
    L_i(\theta, \tau) .
  \end{aligned}
\end{equation}
Note that this is the ``usual'' likelihood in a generalised linear mixed model \cite[cf.][Chapter 7]{demidenko_mixed_2013}. This is analogous to the situation in scheme $\text{FD}$ and $\text{UD}$, where there are no random effects. Studying the behaviour of estimators from this model is, however, more difficult and is the aim of the next section where we will study a simplified version of the random effect schemes.

\section{Estimation in a latent class model}
\label{sec:estim-latent-class}

In the following we consider the non-parametric scenario from $\text{FD}_r$ and $\text{UD}_r$, i.e. the fixed and up-down cases with random effects. We will make the additional simplifying assumption that $\alpha_i$ is binary, i.e. a latent class model with two latent classes $0$ and $A$. The basic model is,
\begin{equation}
  \label{eq:14}
  \begin{aligned}
    & Y_{it} \:\big\vert\: S_{it} = s, \alpha_i \sim
    \text{Bernoulli} \left( \pi_{s} (\alpha_i) \right) \\
    & S_{it} \sim f_{S_{it}} \\
    & \alpha_i = U_i A, \quad A > 0, \quad U_i \sim \text{Bernoulli} \left( \tau \right),
    \quad \tau \in \left( 0, 1 \right) .
\end{aligned}
\end{equation}
Note that $A$ is not a parameter but a known constant signifying the difference between the two latent classes, i.e. if we were willing to assume that the probabilities were linear in the latent class on the psychometric function scale,
\begin{equation}
  \label{eq:5}
  F^{-1}\left(\pi_s(A)\right) - F^{-1}\left(\pi_s(0)\right) = A,
\end{equation}
so that on this scale, $A$ is the difference between the two subjects from the two latent classes evaluated at the same stimulus intensity.

Recall that the stimulus intensity $S$ is assumed to take values in $\setdef{1, \ldots, D}$ and that in the non-parametric scenario we wish to estimate $\theta = \left(\pi_1(0), \ldots, \pi_{D}(0), \pi_1(A), \ldots, \pi_{D}(A) \right)$. Note that the marginal probability of an accurate response at intensity $s$ is given by,
\begin{equation}
  \label{eq:15}
  \begin{aligned}
    \pi_s &= \cproba{Y_{it} = 1}{S_{it} = s} \\
    &= \cproba{\alpha_i = 0}{S_{it} = s} \pi_s (0) + \cproba{\alpha_i = A}{S_{it} = s} \pi_s (A) .
\end{aligned}
\end{equation}
In the fixed design, the marginal probability is the conventional weighted estimator due to independence of $\alpha_i$ and $S_{it}$ under $\text{FD}_r$,
\begin{equation}
  \label{eq:41}
   \pi_s \overset{\text{FD}_r}{=} (1-\tau) \pi_s (0) + \tau \pi_s (A) .
\end{equation}
Following the conditionality arguments also given in Section \ref{sec:estim-under-textfd_r}, we base inference about $\theta$ on the likelihood function from (\ref{eq:53}),
\begin{equation}
  \label{eq:17}
  \begin{aligned}
    L(\theta, \tau)
    &=
    \prod_{i=1}^{N} \int_{}^{}
    f_{\alpha_i} (a_i)
    \prod_{t=1}^{T}
    f_{Y_{it} \:\big\vert\: S_{it}, \alpha_i} \left(y_{it} \:\big\vert\: s_{it}, a_i \right)
    \: d a_i \\
    &=
    \prod_{i=1}^{N} \bigg\{
      (1-\tau)
      \prod_{t=1}^{T}
      \pi_{s_{it}}^{y_{it}} (0) \left(1 - \pi_{s_{it}} (0)\right)^{1- y_{it}} \\
      &\qquad +
      \tau
      \prod_{t=1}^{T}
      \pi_{s_{it}}^{y_{it}} (A) \left(1 - \pi_{s_{it}} (A)\right)^{1-y_{it}}      
      \bigg\} .
  \end{aligned}
\end{equation}
Introducing the total number of trials for subject $i$ at intensity $s$ and the number of correct responses at this intensity, $T_{is} = \sum_{t=1}^{T} \indic{S_{it} = s}$ and $m_{is} = \sum_{t: S_{it} = s} y_{it}$, we obtain,
\begin{equation}
  \label{eq:18}
  \begin{aligned}
    L(\theta, \tau)
    &=
    \prod_{i=1}^{N} \bigg\{
    (1-\tau)
    \prod_{s=1}^{D}
    \pi_{s}^{m_{is}} (0) \left(1 - \pi_{s} (0)\right)^{T_{is} - m_{is}} \\
    &\qquad +
    \tau
    \prod_{s=1}^{D}
    \pi_{s}^{m_{is}} (A) \left(1 - \pi_{s} (A)\right)^{T_{is} - m_{is}}
    \bigg\} .
  \end{aligned}
\end{equation}

We focus, without loss of generality, on estimation of the probabilities $\pi_s(0)$ and $\pi_s(A)$ at intensity $s$. Note that we would in Scheme $\text{FD}_r$ estimate the marginal accuracy probability by the plug-in estimator,
\begin{equation}
  \label{eq:19}
  \widehat{\pi}_s = (1-\widehat{\tau}) \cdot \widehat{\pi}_s (0) +
  \widehat{\tau} \cdot \widehat{\pi}_s (A) .
\end{equation}
Thus, we would need estimates of $\pi_s(0)$, $\pi_s(A)$ (the accuracy probabilities given stimulus level $s$ and latent classes $0$ and $A$), along with $\tau$ (the prevalence of the latent class $A$).

In Appendix \ref{cha:deriv-rand-effect} we derive the likelihood equations for $\theta$ and $\tau$, which do not, to our knowledge, admit any closed-form solutions. To study the properties of the estimator further, we apply expectation maximisation (EM).

\subsection{EM algorithm}
\label{sec:em-algorithm}

As a direct solution to the maximum likelihood equations is intractable, we consider maximum likelihood estimation using an expectation maximisation algorithm \citep{dempster_maximum_1977}. Our purpose is not to suggest an implementation of an EM algorithm, but rather to use the resulting estimates as a basis for discussion of the influence of adaptive designs on small-sample bias. Recall that the basis of the EM algorithm is to construct a lower bound on the likelihood (the E-step), maximise this bound as a function of the parameters (the M-step), and then iterate over this procedure. The bound is made tight around the current parameter estimate by choosing the bound to be the conditional expectation of the random effect given the data evaluated at the current parameter estimates.

More specifically, for a dominating measure $Q_{\alpha}$, the log-likelihood from (\ref{eq:18}) may be written as,
\begin{equation}
  \label{eq:24}
  \begin{aligned}
    l(\theta, \tau)
    &=
    \sum_{i=1}^{N}
    \log \expectund{f_{\bm{Y}_{i} \:\vert\: \bm{S}_{i}, \alpha_i}}{\alpha_i \sim f_{\alpha}} \\
    &=
    \sum_{i=1}^{N}
    \log
    \expectund{
      \frac{f_{\bm{Y}_{i} \:\vert\: \bm{S}_{i}, \alpha_i}}{Q_{\alpha_i}} f_{\alpha_i}
    }{
      \alpha_i \sim Q_{\alpha}
    } \\
    &\geq
    \sum_{i=1}^{N}
    \expectund{
      \log
      \left(
        \frac{f_{\bm{Y}_{i} \:\vert\: \bm{S}_{i}, \alpha_i}}{Q_{\alpha_i}}
        f_{\alpha_i}
      \right)
    }{
      \alpha_i \sim Q_{\alpha}
    } ,    
  \end{aligned}
\end{equation}
where we in the last step applied Jensen's inequality. We take,
\begin{equation}
  \label{eq:25}
  Q_{\alpha_i} (a) = \cproba{\alpha_i = a}{\bm{Y}_i, \bm{S}_i} , \quad \text{for } a = 0, A,
\end{equation}
to be the conditional probability that subject $i$ belongs to the latent class $a$ given the observed data for subject $i$. For this choice of $Q_{\alpha}$ we denote by $\tilde{l}$ the objective function on the right-hand side of (\ref{eq:24}) and notice that this constitutes a lower bound on the likelihood function. We also note that the bound in (\ref{eq:24}) holds for any choice of a distribution $Q$ supported on $\setdef{0, A}$. The particular choice in (\ref{eq:25}) is taken to coincide with the ``usual'' choice of $Q$ in the fixed design case.

The idea is then to obtain weights in the form of the posterior probability that subject $i$ belongs to the latent class $A$ given the data (accuracies and intensities) on subject $i$,
\begin{equation}
  \label{eq:26}
  \begin{aligned}
    w_i &= w_i (\theta, \tau) \\
    &=
    Q_{\alpha_i} (A) \\
    &= \frac{
      f_{\bm{Y}_{i} \:\vert\: \bm{S}_{i}, \alpha_i = A} f_{\bm{S}_{i} \:\vert\: \alpha_i=A} f_{\alpha_i}(A)
    }{
      f_{\bm{Y}_{i} \:\vert\: \bm{S}_{i}, \alpha_i = 0} f_{\bm{S}_{i} \:\vert\: \alpha_i=0} f_{\alpha_i}(0) +
      f_{\bm{Y}_{i} \:\vert\: \bm{S}_{i}, \alpha_i = A} f_{\bm{S}_{i} \:\vert\: \alpha_i=A} f_{\alpha_i}(A)
    } \\
    &=
    \frac{
      1
    }{
      1 +
      \frac{1-\tau}{\tau}
      \exp \left\{l_i(0) - l_i(A) \right\}
      \frac{f_{\bm{S}_{i} \:\vert\: \alpha_i=0}}{f_{\bm{S}_{i} \:\vert\: \alpha_i=A}}
    }
  \end{aligned}
\end{equation}
where $l_i (a)= \log f_{\bm{Y}_{i} \:\vert\: \bm{S}_{i}, \alpha_i = a}$ is the conditional log-likelihood of observing $\bm{Y}_i$ given the intensities $\bm{S}_i$ and the fact that subject $i$ belongs to the latent class $a$. Note that the conditional log-likelihoods are given by,
\begin{equation}
  \label{eq:28}
  l_i (a) = \sum_{s=1}^{D}
  \left\{
    m_{is} \log \pi_s (a) +
    \left(T_{is} - m_{is} \right) \log \left( 1 - \pi_s (a) \right)
  \right\} .
\end{equation}
Further, note that the distribution of the intensities given the latent class $f_{\bm{S}_{i} \:\vert\: \alpha_i}$ is generally complicated. In the fixed design $\text{FD}_r$, however, $\bm{S}_{i} \indep \alpha_i$, so that the ratio in the denominator disappears, i.e.
\begin{equation}
  \label{eq:55}
  w_i
  \overset{\text{FD}_r}{=}
  \frac{
    1
  }{
    1 +
    \frac{1-\tau}{\tau}
    \exp \left\{l_i(0) - l_i(A) \right\}
  } .
\end{equation}
In the $\text{UD}_r$ design the ratio might be estimated by simulation. Assuming that we by some method can calculate the weight, we would then regard the weights as fixed and thus estimate $\theta$ by maximising,
\begin{equation}
  \label{eq:27}
  \begin{aligned}
    \tilde{l}(\theta)
    &=
    \sum_{i=1}^{N}
    \left\{
      w_i \log
      \left(
        \frac{f_{Y_{i} \:\vert\: S_{i}, \alpha_i = A}}{w_i} \tau
      \right) +
      (1 - w_i) \log \left(
        \frac{f_{Y_{i} \:\vert\: S_{i}, \alpha_i = 0}}{1 - w_i} \left[1 - \tau \right]
      \right)
    \right\} \\
    &=
    \sum_{i=1}^{N}
    \left\{
      w_i l_i (A) +
      (1 - w_i) l_i (0)
    \right\} + \text{constant} ,
  \end{aligned}
\end{equation}

The advantage of this method for the present purpose is that the EM algorithm allows for closed form solutions in the M-step. Differentiating the objective function (\ref{eq:27}) with respect to one of the probabilities, say $\pi_s(0)$, we see that
\begin{equation}
  \label{eq:30}
  \frac{\partial}{\partial \pi_s(0)}
  \tilde{l} (\theta)
  =
  \sum_{i=1}^{N} (1-w_i)
  \left\{
    \frac{m_{is}}{\pi_s(0)} - \frac{T_{is} - m_{is}}{1-\pi_s(0)}
  \right\} ,
\end{equation}
so that the estimation equation $\partial \tilde{l}(\theta) / \partial \pi_s(0) = 0$ is satisfied for,
\begin{equation}
  \label{eq:31}
  \hat{\pi}_s(0) = \frac{
    \sum_{i=1}^{N} \left(1 - w_i\right) m_{is}
  }{
    \sum_{i=1}^{N} \left(1 - w_i\right) T_{is}
  } .
\end{equation}
Analogously, we obtain,
\begin{equation}
  \label{eq:32}
  \hat{\pi}_s(A) = \frac{
    \sum_{i=1}^{N} w_i m_{is}
  }{
    \sum_{i=1}^{N} w_i T_{is}
  } .
\end{equation}
We also note that once the weights $\setdef{w_i}_{i=1}^N$ and the probabilities $\theta$ are known, we may heuristically estimate $\tau$ by, 
\begin{equation}
  \label{eq:33}
  \hat{\tau} = \frac{1}{1 + \hat{\chi}} ,
\end{equation}
where
\begin{equation}
  \label{eq:34}
  \chi = \frac{
    \sum_{i=1}^{N} \left(1- w_i \right) / w_i
  }{
    \sum_{i=1}^{N} \exp \left\{ l_i(0) - l_i(A) \right\}
    \frac{f_{S_{i} \:\vert\: \alpha_i=0}}{f_{S_{i} \:\vert\: \alpha_i=A}}
  } .
\end{equation}
Indeed this follows simply by taking the reciprocal expression for the weights in (\ref{eq:26}), summing and solving for $\tau$. Note that this results in an estimator $\hat{\tau}$ which is strictly positive. 

In summary, the EM algorithm may be described as follows. We initialise by choosing some $\theta^{(0)}$ and $\tau^{(0)}$ and then proceed iteratively with the following steps.
\begin{enumerate}
\item{E-step:} Obtain the $(r+1)$'st weights by
  \begin{equation}
    \label{eq:29}
    w^{(r+1)}_i = w_i (\theta^{(r)}, \tau^{(r)}) ,
  \end{equation}
  using formula (\ref{eq:26}).
\item{M-step:} Calculate the estimates,
  \begin{equation}
    \label{eq:35}
    \hat{\pi}^{(r+1)}_s(0) = \frac{
      \sum_{i=1}^{N} \left(1 - w^{(r+1)}_i\right) m_{is}
    }{
      \sum_{i=1}^{N} \left(1 - w^{(r+1)}_i\right) T_{is}
    } ,
  \end{equation}
  and
  \begin{equation}
    \label{eq:36}
    \hat{\pi}^{(r+1)}_s(A) = \frac{
      \sum_{i=1}^{N} w^{(r+1)}_i m_{is}
    }{
      \sum_{i=1}^{N} w^{(r+1)}_i T_{is}
    } .
  \end{equation}
  Also, calculate $\tau^{(r+1)}$ by plugging $\setdef{w_i^{(r+1)}}_{i=1}^N$ and $\theta^{(r+1)}$ into the formulas (\ref{eq:33})-(\ref{eq:34}).
\end{enumerate}
Using \mbox{$\absval{(\theta^{(r+1)}, \tau^{(r+1)}) - (\theta^{(r)}, \tau^{(r)})}$} to assess convergence following each iteration, the algorithm is stopped once this difference becomes sufficiently small. Supposing that this happens after $R$ iterations, we take $(\hat{\theta}, \hat{\tau}) = (\theta^{(R)}, \tau^{(R)})$.

Comparing to the likelihood equations in Appendix \ref{cha:deriv-rand-effect}, the EM-algorithm may be viewed as an iterative solution in which we switch between solving the equations (\ref{eq:21})-(\ref{eq:22}) and solving (\ref{eq:23}), in each step sharing current estimates between equations.

\subsection{Bias of the EM estimates}
\label{sec:bias-em-estimates}

We consider now in more detail the small sample properties of the estimator resulting from the EM algorithm. We will derive such properties under the na\"{i}ve presumption that the weight
\begin{equation}
  \label{eq:46}
  w_i = \cprobaund{\alpha_i = A}{Y_i, S_i}{(\theta, \tau)} ,
\end{equation}
is known, while in practise, the weight must be estimated so that the probability in (\ref{eq:46}) is evaluated at the estimates $(\hat{\theta}, \hat{\tau})$. We still perceive the weight as stochastic as a function of the data but ignore the source of randomness stemming from estimation of the parameters. Assuming that we did not need to apply parameter estimates in the weight, consider the estimator,
\begin{equation}
  \label{eq:44}
  \check{\pi}_s(A) = \frac{
    \sum_{i=1}^{N} w_i m_{is}
  }{
    \sum_{i=1}^{N} w_i T_{is}
  } ,
\end{equation}
where we use a ``check'' rather than a ``hat'' to differentiate from the estimator using the estimated weights. We begin by calculating,
\begin{equation}
  \label{eq:45}
  \begin{aligned}
    &\expect{\check{\pi}_s (A) \sum_{i=1}^{N} w_i T_{is}} -
    \pi_s(A) \expect{\sum_{i=1}^{N} w_i T_{is}} \\
    &=
    \sum_{i=1}^{N} \expect{w_i m_{is}} -
    \pi_s(A) \sum_{i=1}^{N} \expect{\cexpect{\indic{\alpha_i = A}}{Y_i, S_i} T_{is}} \\
    &=
    \sum_{i=1}^{N} \expect{\cexpect{\indic{\alpha_i = A} m_{is}}{Y_i, S_i}} -
    \pi_s(A) \sum_{i=1}^{N} \expect{\cexpect{\indic{\alpha_i = A}  T_{is}}{Y_i, S_i}} \\
    &=
    \sum_{i=1}^{N} \expect{\indic{\alpha_i = A} m_{is}} -
    \pi_s(A) \sum_{i=1}^{N} \expect{\indic{\alpha_i = A}  T_{is}} ,
  \end{aligned}
\end{equation}
noting that $T_{is}$ and $m_{is}$ are functions of $S_i$ and $(Y_i,S_i)$, respectively. Continuing by inserting the definition of $m_{is}$ and $T_{is}$,
\begin{equation}
  \label{eq:42}
  \begin{aligned}
    &=
    \sum_{i=1}^{N} \expect{\sum_{t=1}^{T} \indic{\alpha_i = A} \indic{S_{it} = s} Y_{it}} -
    \pi_s(A) \sum_{i=1}^{N} \expect{\sum_{t=1}^{T} \indic{\alpha_i = A}  \indic{S_{it} = s}} \\
    &=
    \sum_{i=1}^{N} \sum_{t=1}^{T}
    \cexpect{Y_{it}}{\alpha_i = A, S_{it} = s} \proba{\alpha_i = A, S_{it} = s} \\
    &\qquad-
    \pi_s(A) \sum_{i=1}^{N} \sum_{t=1}^{T} \proba{\alpha_i = A, S_{it} = s} \\
    &=
    \pi_s(A)
    \sum_{i=1}^{N} \sum_{t=1}^{T}
    \proba{\alpha_i = A, S_{it} = s} \\
    &\qquad -
    \pi_s(A) \sum_{i=1}^{N} \sum_{t=1}^{T} \proba{\alpha_i = A, S_{it} = s}  \\
    &=
    0 .
  \end{aligned}
\end{equation}
This means that
\begin{equation}
  \label{eq:47}
  \begin{aligned}
    0 &=
    \expect{\check{\pi}_s (A) \sum_{i=1}^{N} w_i T_{is}} -  \pi_s(A) \expect{\sum_{i=1}^{N} w_i T_{is}} \\
    &=
    \covari{\sum_{i=1}^{N} w_i T_{is}}{\check{\pi}_s (A)} +
    \left(\expect{\check{\pi}_s(A)} -  \pi_s(A)\right) \expect{\sum_{i=1}^{N} w_i T_{is}} ,
  \end{aligned}
\end{equation}
so that rearranging we arrive at the formula,
\begin{equation}
  \label{eq:48}
  \text{Bias}(\check{\pi}_s(A)) = - \frac{
    \covari{\sum_{i=1}^{N} w_i T_{is}}{\check{\pi}_s (A)}
  }{
    \expect{\sum_{i=1}^{N} w_i T_{is}} 
  } .
\end{equation}
We see that this is essentially the same as formula (\ref{eq:13}) with the total replaced by the weighted total.

Examining formula (\ref{eq:48}), we observe that it coincides with the formula for the scenario with no random effects if the weights are constant (i.e. $w_i \equiv constant$ for all $i=1,\ldots,N$) and in particular if $\tau = 1$, cf. the formula for the weights in (\ref{eq:26}), corresponding to a scenario where all individuals belong to the latent class $A$. To further elucidate the difference between the scenarios with and without random effects it is illustrative to decompose the covariance in the numerator of the bias formula (\ref{eq:48}),
\begin{equation}
  \label{eq:43}
  \begin{aligned}
    &\covari{\sum_{i=1}^{N} w_i T_{is}}{\check{\pi}_s (A)} \\
    &=
    \sum_{i=1}^{N}
    \left\{
      \covari{ \cexpect{w_i T_{is}}{\bm{S}_i}}{\cexpect{\check{\pi}_s (A)}{\bm{S}}}
      +
      \expect{T_{is} \ccovari{w_i}{\check{\pi}_s (A)}{\bm{S}}}
    \right\}
    ,
  \end{aligned}
\end{equation}
by the Law of Total Covariance. We see that in the case of no random effect the second, conditional, covariance is zero as the weights are constant, while in the case with random effects, there are two covariances that contribute to the bias expression: The covariance between the conditional expectation of the estimated accuracy at intensity $s$ and the conditional weighted total samples at this intensity along with the conditional covariance between the estimator and the weights. The latter covariance signifies the relationship between the estimator and the posterior probability of belonging to class $A$. Intuitively, we would expect this covariance to be positive if the latent class $A$ consists of those with high accuracy, so that a higher estimate of the accuracy was associated with a higher probability of belonging to the latent class $A$. The signs of the two covariances interact to magnify or attenuate the bias when comparing to the scenario with no random effects, but without introducing further assumptions, it is difficult to compare the scenarios with and without random effects and we do not pursue this further.

\section{A simulation study}
\label{sec:simulations}

Below we present a simulation study, which compares the performance of estimators in the parametric setup. Specifically, we simulate data from the four schemes $\text{FD}$, $\text{UD}$, $\text{FD}_r$ and $\text{UD}_r$ and compare estimators for the intercept and slope of the logistic psychometric function in (\ref{eq:6}) while varying the number of participants $N$ and the number of trials per participant $T$.

\subsection{Methods}
\label{sec:methods}

In the simulation study, data from four schemes are simulated and analysed: $\text{FD}$ (fixed design, no random effect), $\text{FD}_r$ (fixed design, random effect), $\text{UD}$ (Up-down design, no random effect), $\text{UD}_r$ (Up-down design, random effect). Each scheme is simulated for a number of parameter setups, and we refer to the combination of a scheme and a parameter setup as a ``scenario''.

A single replication of a simulation scenario may be summarised as follows,
\begin{enumerate}
\item Simulate data \textsf{df} from a function \textsf{simulFun(params)} for a given set of parameters \textsf{params}.
\item Analyse the data using the function \textsf{analysisFun(data = df)}.
\item Obtain and store estimates and their associated standard errors.
\end{enumerate}
The relevant functions for the four schemes are given in Table \ref{tab:sim-schemes}.
Each scenario is based on $R = 1000$ Monte Carlo replications. 

\begin{table}[htb]
  \centering
  \begin{tabularx}{0.7\linewidth}{l | c c | c c}
    Scheme & Design & Random intercept & \textsf{simulFun} & \textsf{analysisFun} \\
    \hline
    $\text{FD}$ & Fixed & No & \textsf{SimFD} & \textsf{glm()} \\
    $\text{FD}_r$ & Fixed & Yes & \textsf{SimFD\_r} & \textsf{glmer()} \\
    $\text{UD}$ & Up-down & No & \textsf{SimUD} & \textsf{glm()} \\
    $\text{UD}_r$ & Up-down & Yes & \textsf{SimUD\_r} & \textsf{glmer()}
  \end{tabularx}
  \caption{Summary of simulation schemes.}
  \label{tab:sim-schemes}
\end{table}

The simulation requires implementation of the functions \textsf{SimFD} and \textsf{SimUD} (where the corresponding \textsf{\_r} functions are obtained by including a random effect), which is done by coding the two sample designs described in Section \ref{sec:two-depend-schem}. We simulate under the parametric setup with logistic psychometric functions. The stimulus intensities take values in $\mathcal{S} = \setdef{d/L, 2d/L, \ldots, d}$ with $\absval{\mathcal{S}} = L$, where $L$ and $d$ are the number of stimulus intensities and maximal intensity, respectively, to be set in the parameters \textsf{params}. This corresponds to the previous setup, but rather than integer intensity steps, the stimulus intensities are updated in steps of size $d/L$. In all cases, the initial intensity $S_1$ was chosen by uniform random sampling from $\mathcal{S}$.

Parameters \textsf{params} were chosen from the setups in Table \ref{tab:parameter-setups}. We fix the parameters of the psychometric function and focus on changing the size of the experiment. We see that the combinations of the four schemes with the 12 setups leads to a total of 48 scenarios. In each scenario, 1000 data sets were simulated. In addition to the parameters in Table \ref{tab:parameter-setups}, we also consider $ED50 = -a/b$, i.e. the stimulus intensity associated with a fifty percent success probability, in the random effects scenario for an average individual (one for whom the random intercept is zero). 

\begin{table}[htb]
  \centering
  \begin{tabularx}{0.5\linewidth}{l | rrrrrrr}
    \multicolumn{1}{c}{Setup} &
                                \multicolumn{1}{c}{$N$} &
                                                          \multicolumn{1}{c}{$T$} &
                                                                                    \multicolumn{1}{c}{$d$} &
                                                                                                              \multicolumn{1}{c}{$L$} &
                                                                                                                                        \multicolumn{1}{c}{$a$} &
                                                                                                                                                                  \multicolumn{1}{c}{$b$} &
                                                                                                                                                                                            \multicolumn{1}{c}{$\tau$}\\
    \hline
    $ 1$&$ 25$&$ 25$& \multirow{4}{*}{$0.2$} & \multirow{4}{*}{$10$} & \multirow{4}{*}{$0.05$} & \multirow{4}{*}{$9$} & \multirow{4}{*}{$1$}\\
    $ 2$&$ 25$&$ 50$&&&&&\\
    $ 3$&$ 25$&$ 75$&&&&&\\
    $ 4$&$ 25$&$100$&&&&&\\
    \hline
    $ 5$&$ 50$&$ 25$& \multirow{4}{*}{$0.2$} & \multirow{4}{*}{$10$} & \multirow{4}{*}{$0.05$} & \multirow{4}{*}{$9$} & \multirow{4}{*}{$1$}\\
    $ 6$&$ 50$&$ 50$&&&&&\\
    $ 7$&$ 50$&$ 75$&&&&&\\
    $ 8$&$ 50$&$100$&&&&&\\
    \hline
    $ 9$&$100$&$ 25$& \multirow{4}{*}{$0.2$} & \multirow{4}{*}{$10$} & \multirow{4}{*}{$0.05$} & \multirow{4}{*}{$9$} & \multirow{4}{*}{$1$}\\
    $10$&$100$&$ 50$&&&&&\\
    $11$&$100$&$ 75$&&&&&\\
    $12$&$100$&$100$&&&&&
  \end{tabularx}
  \caption[Parameter setups in the simulation study.]{Parameter setups in the simulation study. $N$ is the number of subject, $T$ is the number of replicated per subject. $L$ is the number of stimulus intensity levels, $d$ is the maximum stimulus level. $a$ is the intercept parameter, while $b$ is the slope of the logistic psychometric function. The between subject heterogeneity is denoted $\tau$.}
  \label{tab:parameter-setups}
\end{table}
Simulations were performed using \textsf{R} (version 4.1.3). Random intercept logistic models were fitted using the package \textsf{lme4} (version 1.1-28).

\subsection{Results}
\label{sec:results}

Below we present results from the simulations using different summary measures. For a choice of parameter $\omega$ and simulated estimates $\setdef{\hat{\omega}_r}$ we calculated the absolute bias,
\begin{equation}
  \label{eq:38}
  \text{absBias}(\hat{\omega}) = \frac{1}{R}\sum_{r=1}^{R}\hat{\omega}_r - \omega
  = \bar{\omega}_{.} - \omega,
\end{equation}
the relative bias,
\begin{equation}
  \label{eq:37}
  \text{relBias }(\hat{\omega})= \frac{\text{absBias} (\hat{\omega})}{\omega} ,
\end{equation}
and the standard error,
\begin{equation}
  \label{eq:39}
  \text{SE}(\hat{\omega})
  = \frac{1}{R-1} \sum_{r=1}^{R} \left(\hat{\omega}_r - \bar{\omega}_{.}\right)^2 ,
\end{equation}
which was calculated as the empirical standard deviation of the estimates. We additionally calculated the root mean square error (RMSE),
\begin{equation}
  \label{eq:40}
  \text{RMSE}(\hat{\omega})
  = \sqrt{ \text{SE}^2(\hat{\omega}) + \text{absBias}^2(\hat{\omega}) } .
\end{equation}
Below, we focus on comparison of fixed and adaptive schemes, i.e. comparing $\text{FD}$ to $\text{UD}$ and $\text{FD}_r$ to $\text{UD}_r$.

\subsubsection{Comparison of schemes}
\label{sec:comparison-schemes}

\begin{figure}[phtb]
  \centering
  \includegraphics[width = 0.49\linewidth]{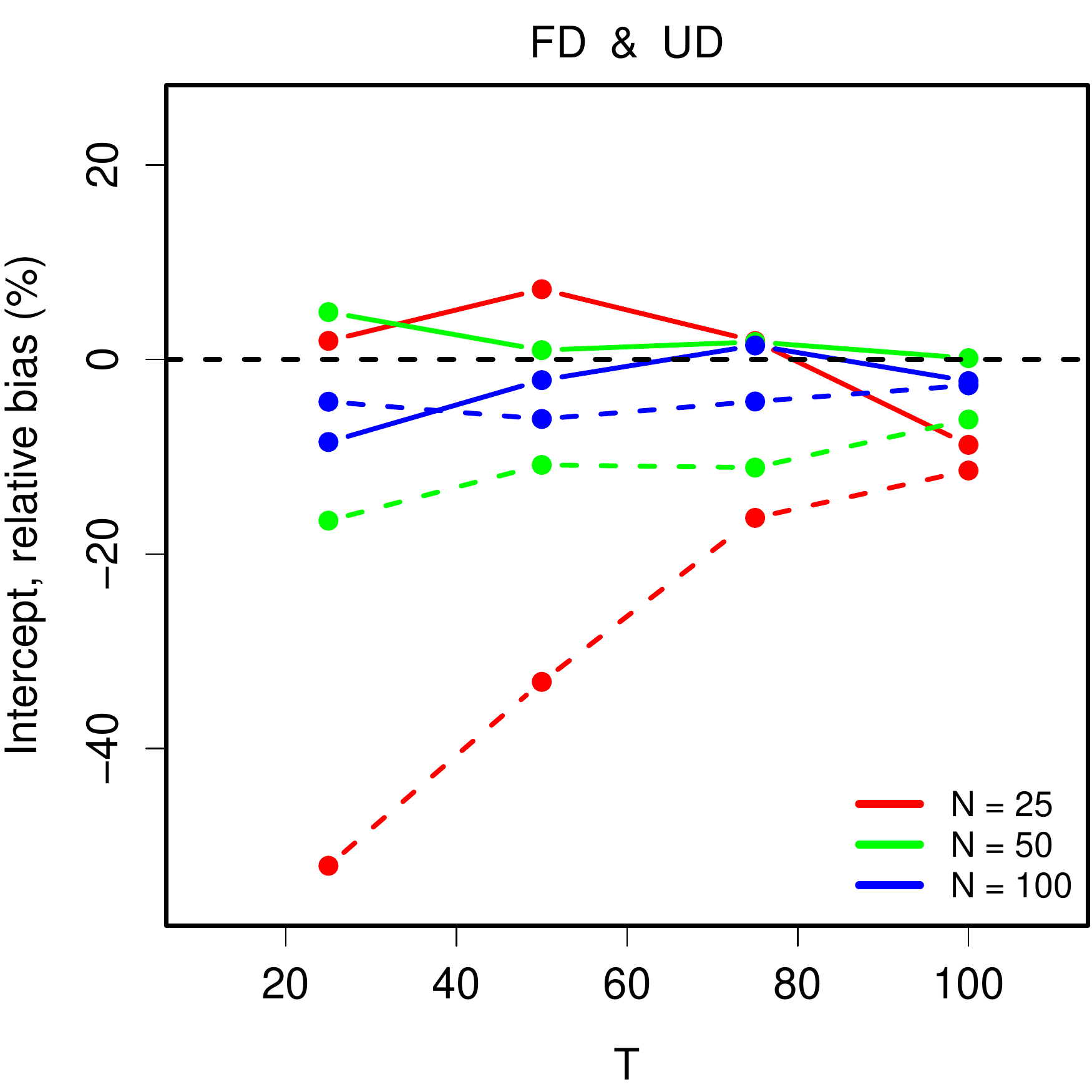}
  \includegraphics[width = 0.49\linewidth]{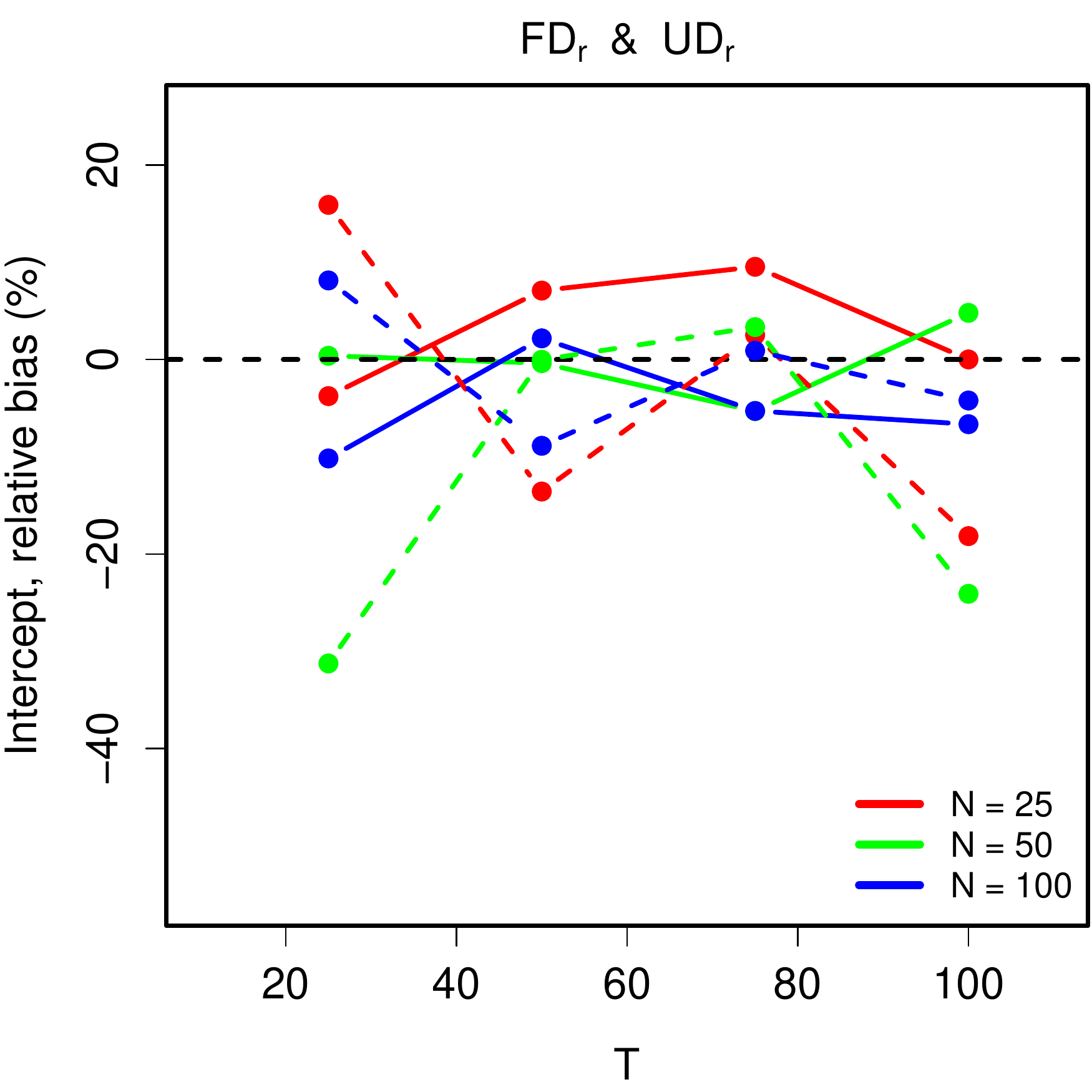}
  \includegraphics[width = 0.49\linewidth]{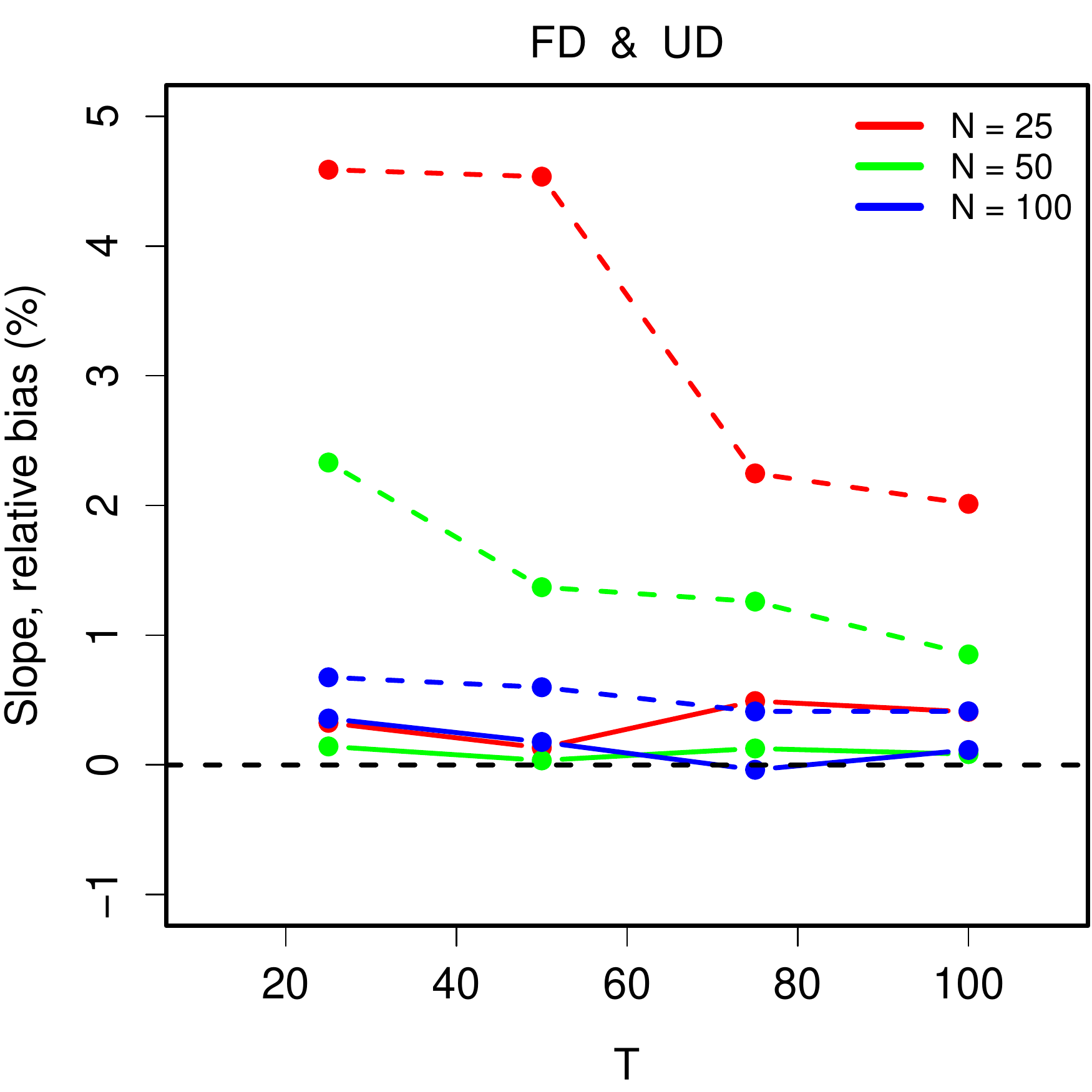}
  \includegraphics[width = 0.49\linewidth]{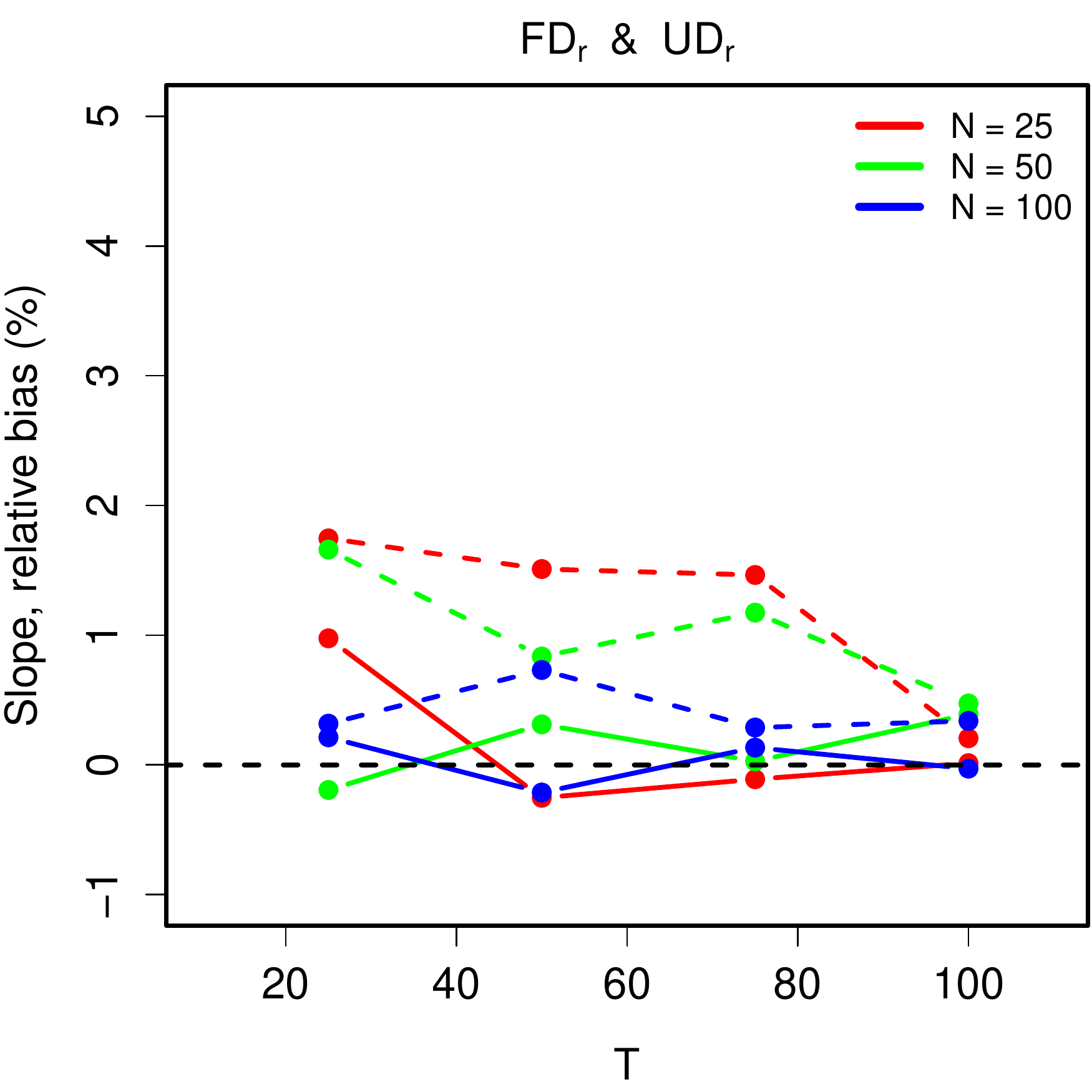}
  \includegraphics[width = 0.49\linewidth]{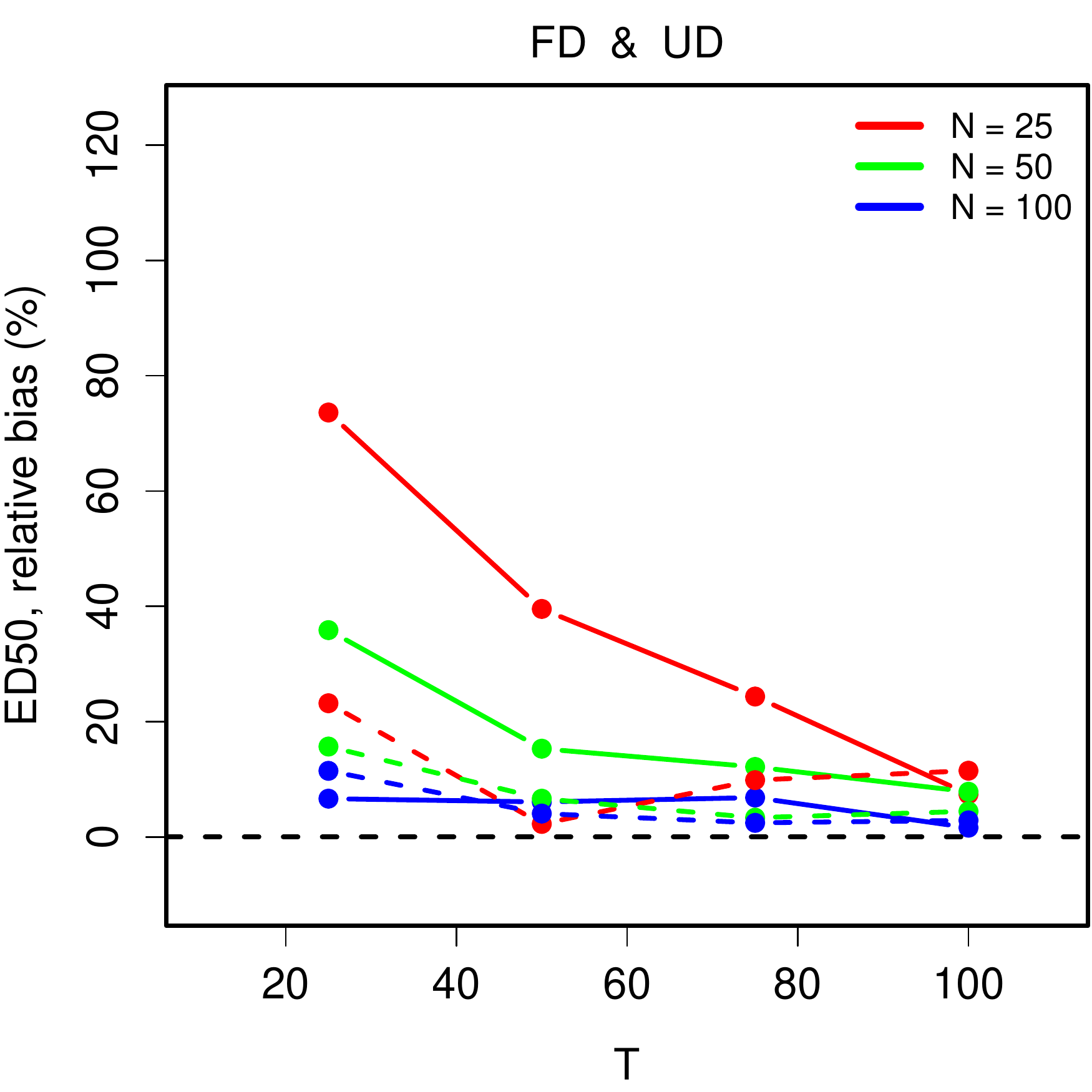}
  \includegraphics[width = 0.49\linewidth]{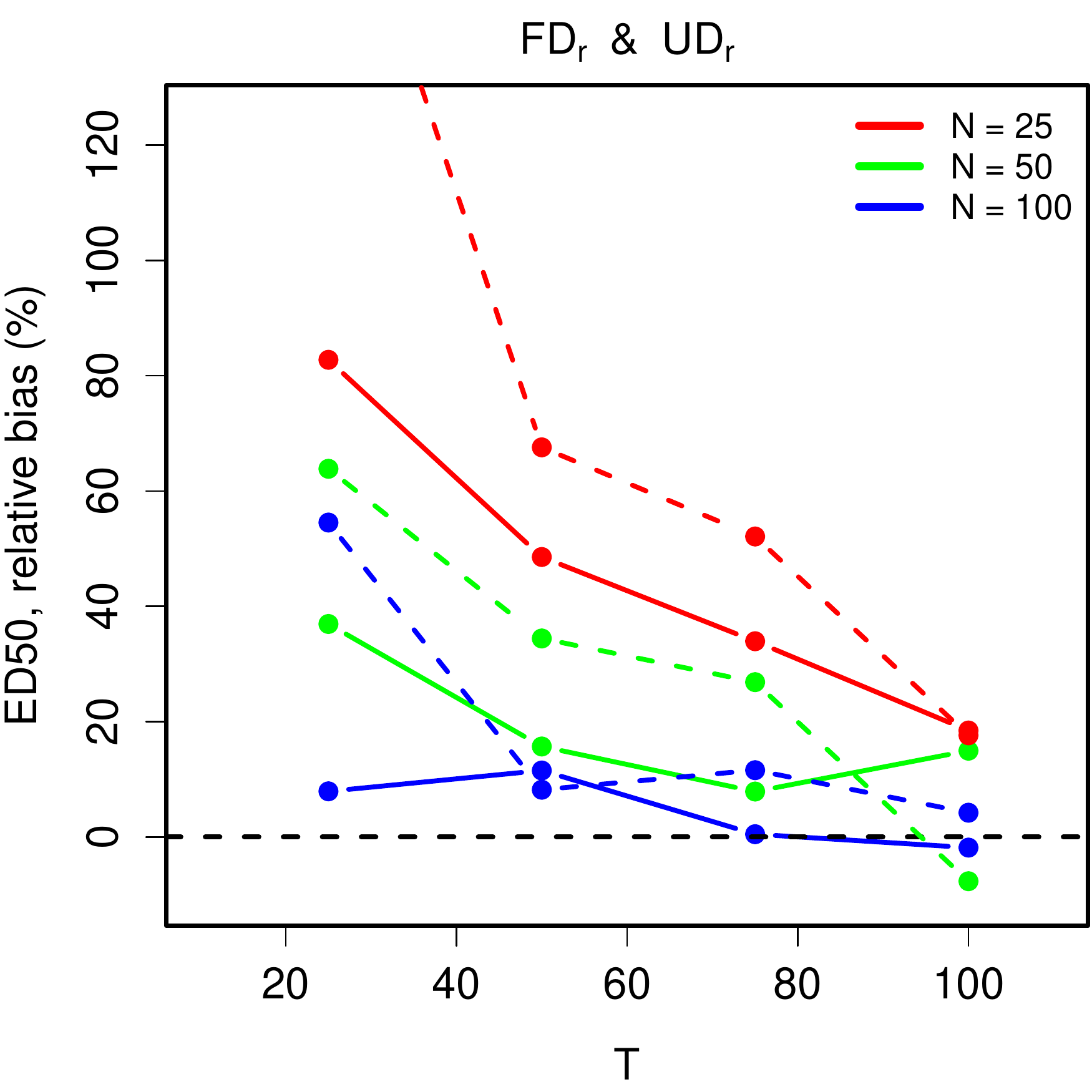}
  \caption[Simulation results for relative bias.]{Simulation results for relative bias on the y-axis and number of trials $T$ per subject on the x-axis. In each plot the full line is the fixed design while the dashed is the up-down design, one line for each choice of number of subjects $N$. The top row is for the intercept parameter, middle is the slope $b$, while the bottom is the ED50. The left column is the scheme with no random effect, while the right column is schemes with a random effect.}
  \label{fig:sim-res-compar}
\end{figure}

\begin{figure}[phtb]
  \centering
  \includegraphics[width = 0.49\linewidth]{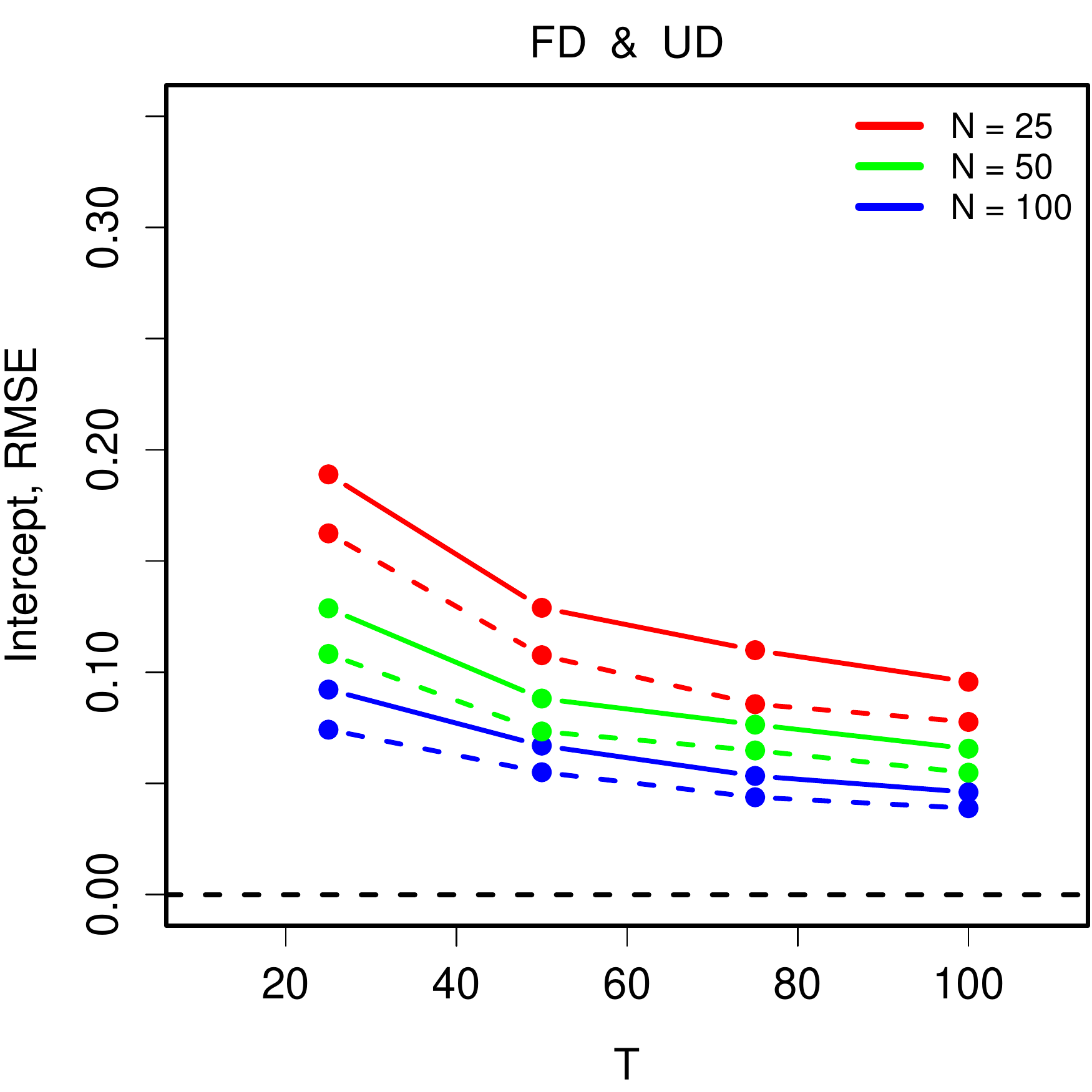}
  \includegraphics[width = 0.49\linewidth]{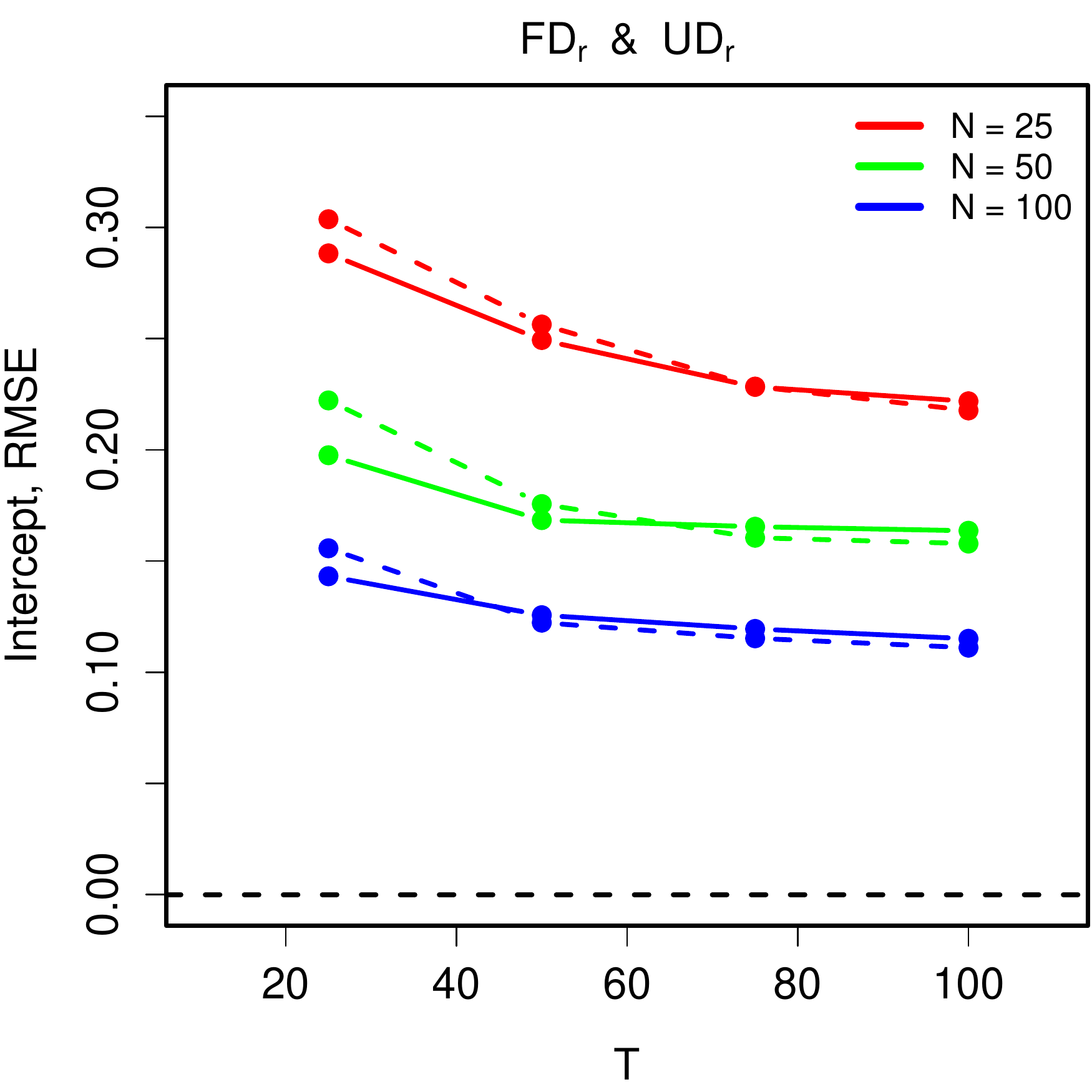}
  \includegraphics[width = 0.49\linewidth]{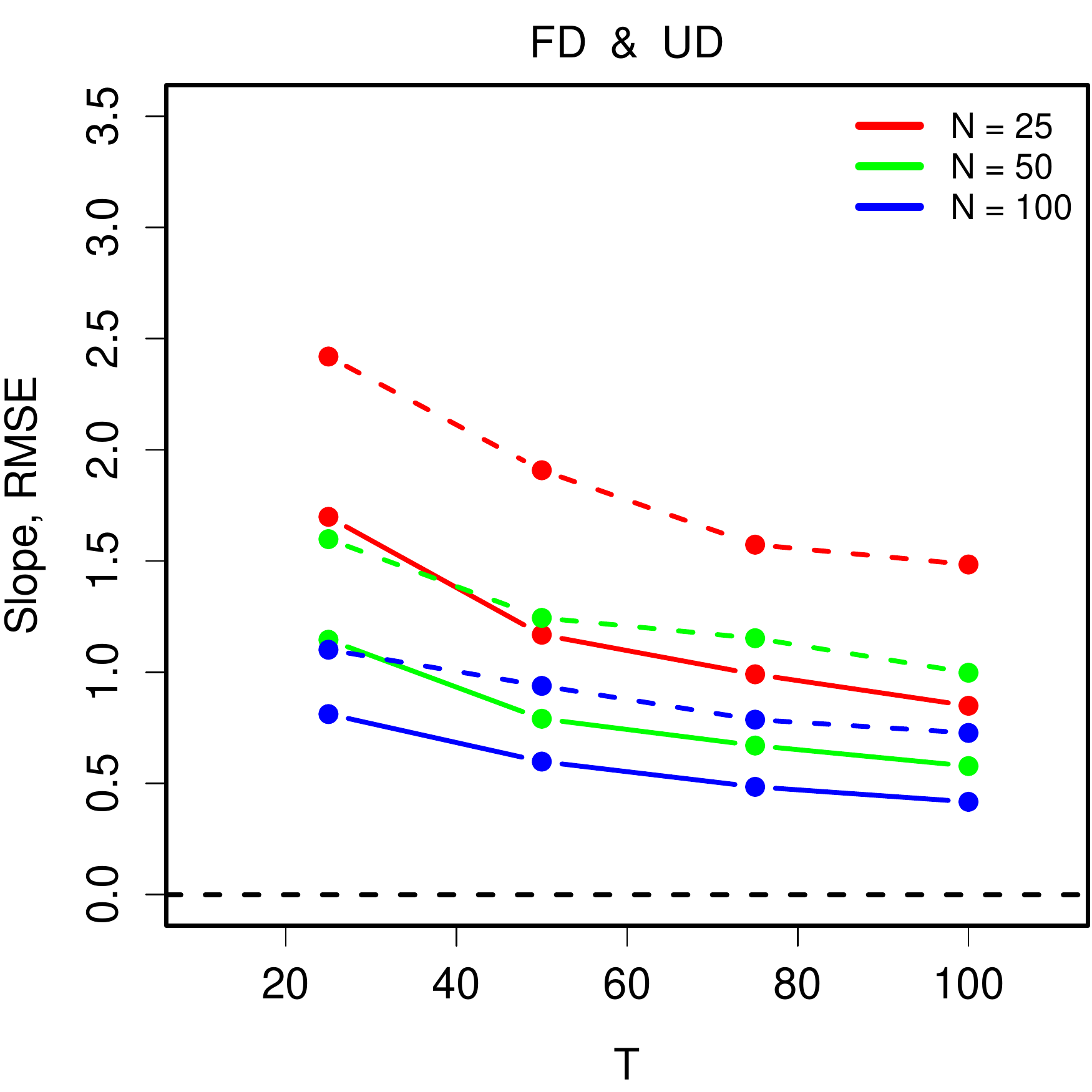}
  \includegraphics[width = 0.49\linewidth]{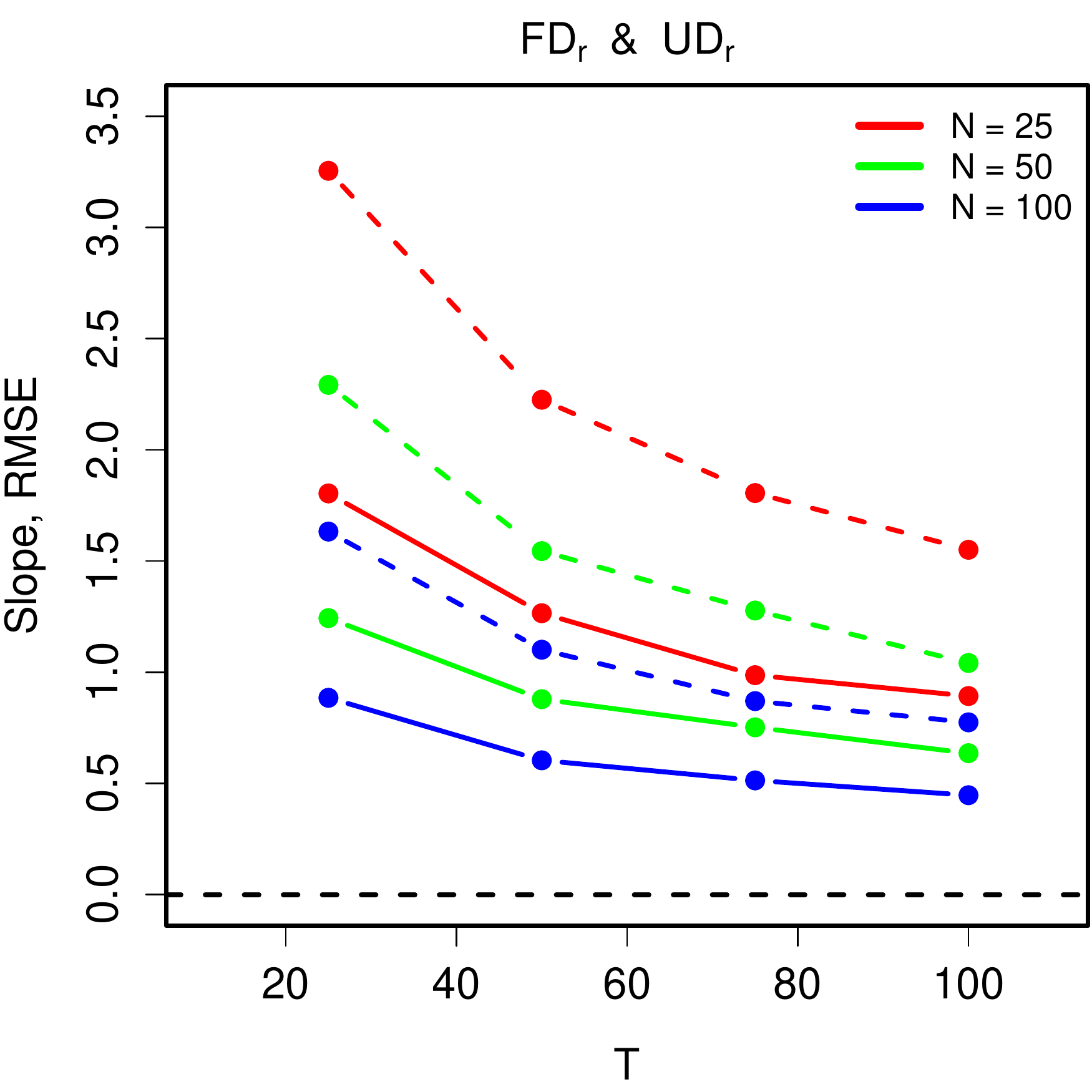}
  \includegraphics[width = 0.49\linewidth]{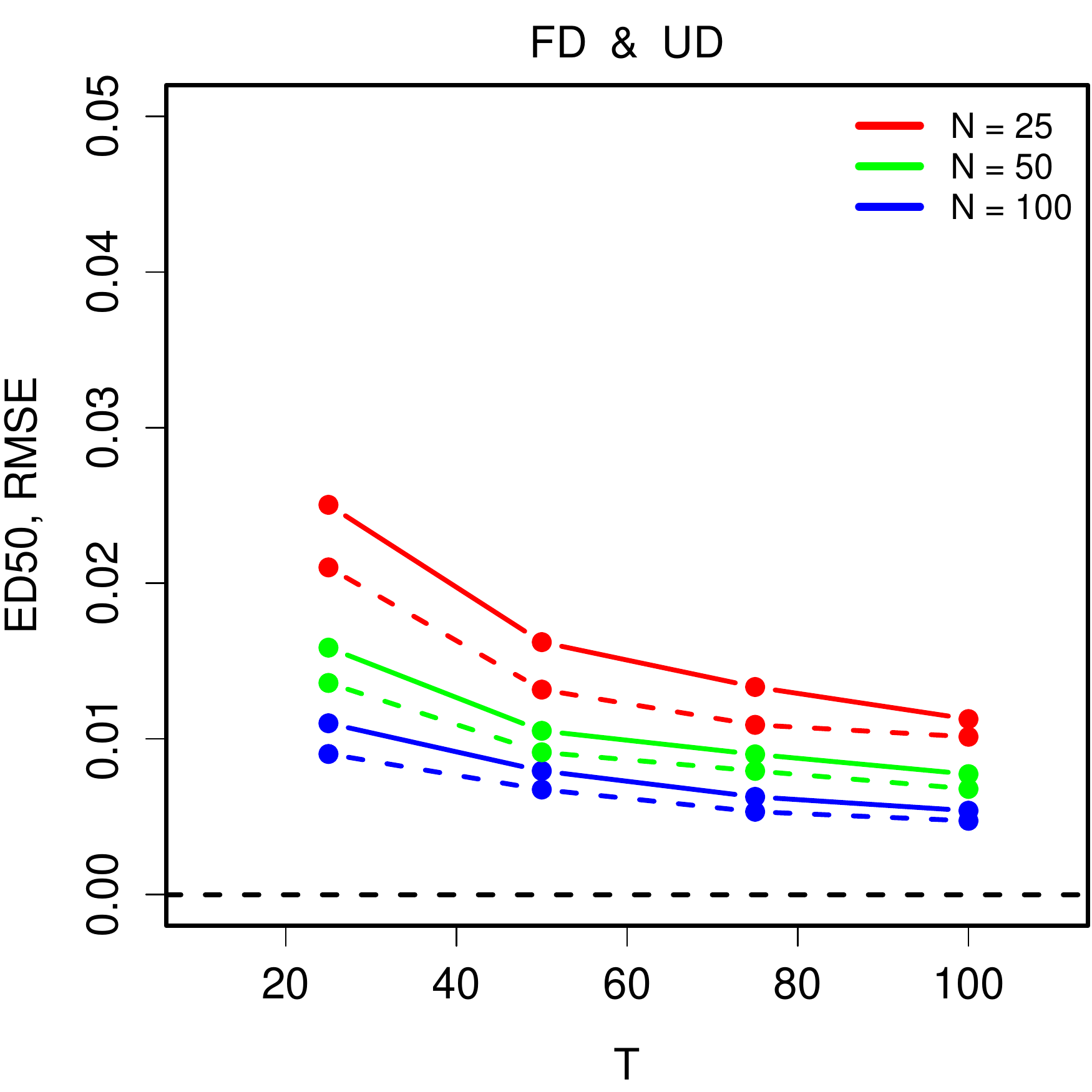}
  \includegraphics[width = 0.49\linewidth]{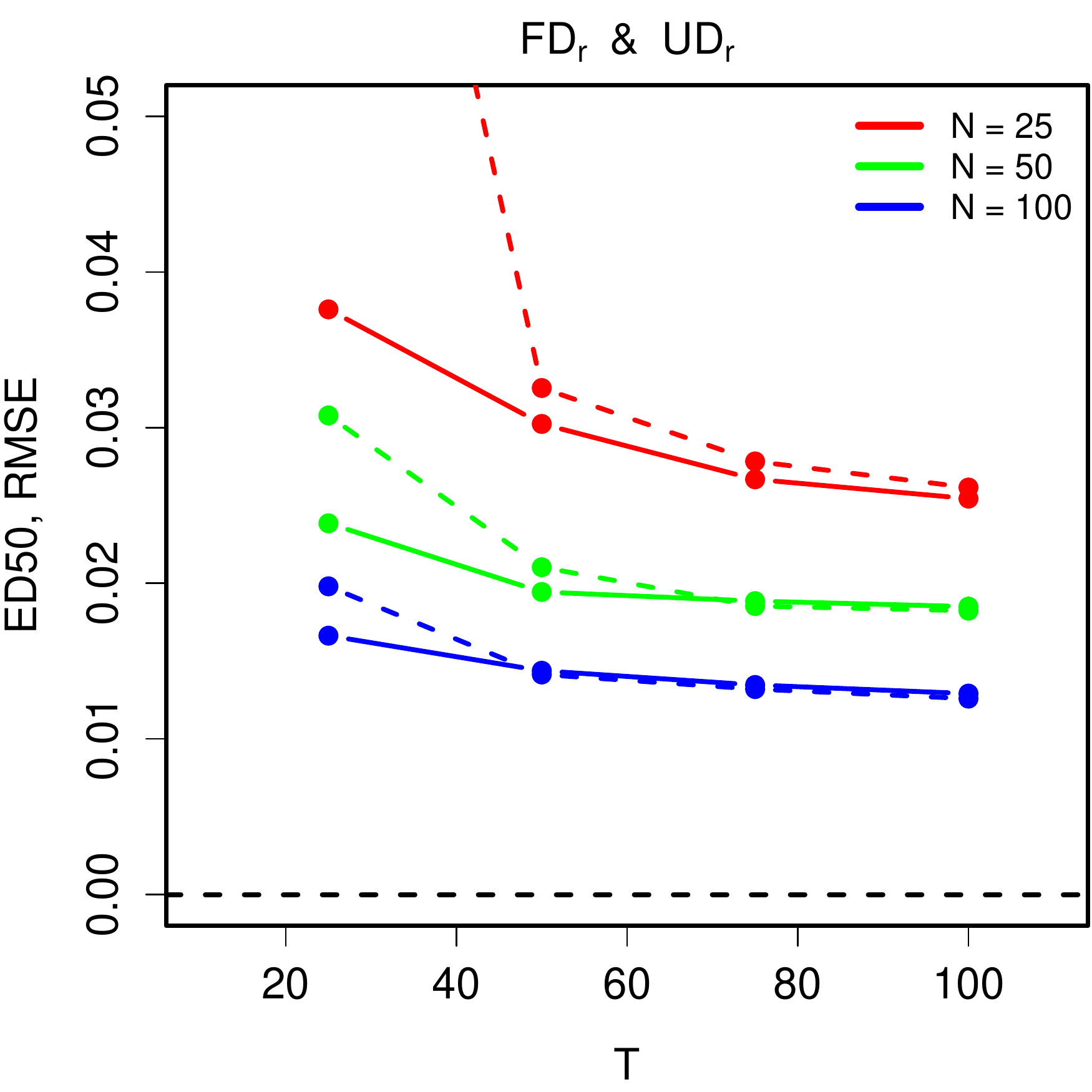}
  \caption[Simulation results for root mean square error.]{Simulation results for root mean square error on the y-axis and number of trials $T$ per subject on the x-axis. In each plot the full line is the fixed design while the dashed is the up-down design, one line for each choice of number of subjects $N$. The top row is for the intercept parameter, middle is the slope $b$, while the bottom is the ED50. The left column is the scheme with no random effect, while the right column is schemes with a random effect.}
  \label{fig:sim-res-compar-rmse}
\end{figure}

Figure \ref{fig:sim-res-compar} depicts the relative bias, while the root mean square error is shown in Figure \ref{fig:sim-res-compar-rmse}. When comparing the fixed design (scheme $\text{FD}$ and $\text{FD}_r$) to the up-down design (scheme $\text{UD}$ and $\text{UD}_r$) we make the following observations, 
\begin{enumerate}
\item The magnitude of the relative bias in the intercept is increased for the up-down design compared to the fixed-design when there is no random effect (scheme $\text{FD}$ and scheme $\text{UD}$).
\item The bias in the intercept is more or less comparable between the two schemes when there is a random effect (scheme $\text{FD}_r$ and scheme $\text{UD}_r$). Indeed, for the case with small number of subjects $N=25$, the bias seems higher for the fixed design, although recall that the true value of $a$ is small so that the relative bias is quite high.
\item With regards to the slope, the bias is higher for the up-down designs, although less inflated when there is a random effect (scheme $\text{FD}_r$ and scheme $\text{UD}_r$). For a high number of trials ($T=100$) and a intermediate to high number of subjects ($N \geq 50$), the bias is smaller.
\item Looking at the RMSE, the up-down design seems better at estimating intercepts, offering a smaller error when there is no random effect and a slightly larger error when there is a random effect present.
\item The RMSE for the slope tells the opposite story: Here the fixed design is better, more so when a random effect is present. Again, this holds especially when sample sizes are small.
\item For estimating the $ED50$, the up-down design is superior in terms of relative bias and slightly superior in RMSE for scenarios with no random effects, while the fixed and up-down design seem comparable when random effects are present.
\end{enumerate}

\section{Discussion}
\label{sec:discussion}

The problem of bias in the slope following from the up-down design as observed in the simulation study is well known in the psychophysics literature. \cite{kaernbach_slope_2001} describes the slope bias as a consequence of the sequential dependence between the observations under an adaptive design, and further demonstrates that the bias persists also under other adaptive designs including stochastic approximation schemes. The bias could also be explained from our interpretation of the bias formula in (\ref{eq:13}): Accuracy probabilities towards the upper edge of the psychometric function are at risk of being overestimated, as trials where an early overestimate occurs might not be corrected as the adaptive design will migrate towards lower stimulus intensities. Similarly, an adaptive design will generally lead to low accuracies being underestimated. Thus, the psychometric function is wrung out of shape by the upper and lower edges leading to the upwards bias of the slope observed in Figure \ref{fig:sim-res-compar}. This would also serve to explain why the intercept estimate, lying close to the middle of the psychometric function in our simulation setup, is less affected. An extension of this argument also explains why the $ED50$ is not affected as it is by definition in the middle of the psychometric function. Of course, we also note the heuristic nature of these interpretations, in that the slope bias refers to the parametric scenario while the bias formula (\ref{eq:13}) describes the non-parametric scenario. We also stress that this bias is a small-sample problem and that the properties of the estimators are preserved asymptotically as argued in Section \ref{sec:infer-fixed-adapt}. This is reflected also in the simulations by the diminishing bias for increased $N$ and $T$.

The denominator of the bias formulae (\ref{eq:13}) and (\ref{eq:48}) along with the presented simulations also serve to show that the bias attenuates fairly quickly as the number of participants and trials increases. However, this should be interpreted cautiously, as one could also imagine realistic scenarios which might interfere with this attenuation. For instance, the presence of learning effects and fatigue would lead to a rise in accuracies in the beginning of the trial which then decreases towards the trial's end. Such effects would render the true accuracy probability a moving target with which the adaptive design might struggle to keep up and thus hindering the convergence observed in Figures \ref{fig:sim-res-compar} and \ref{fig:sim-res-compar-rmse}. A more precise characterisation of the convergence under such effects could be investigated through further simulation.

In most psychophysical trials, the number of participants is rather low, but it is often possible to collect a fairly high number of trials, as a single trial does not demand much time and effort from a participant. In some psychometric studies, the task to be solved in a single trial may be more involved, for instance if a participant must read a snippet of text and choose a response from a list of alternatives. In these cases, the amount of effort and time spent on a single task will curtail the number of trials to be low. Further, we have not considered the additional complication connected to applying data-driven stopping, which would make the number of trials $T$ a random variable.

The objective when choosing to apply an adaptive design is to make the sampling more efficient. As we have shown, however, the adaptive allocation will exert itself also in the opposite direction by requiring more samples to be taken to balance out the adverse small-sample properties implied by the adaptation. This begs the need for careful consideration and performance of simulations studies prior to choosing the design.

\section*{Acknowledgements}

SBK visited Hasselt University and the Center for Statistics, Belgium, in the autumn of 2019 and would like to thank Hasselt University for their hospitality. The authors are grateful to Professor Geert Molenberghs for many interesting discussions in the early phases of this manuscript. 

\medskip

\renewcommand{\bibname}{References}

\appendix

\chapter{Derivations for random effect schemes}
\label{cha:deriv-rand-effect}

\section{Maximum likelihood estimation}
\label{sec:maxim-likel-estim}

We obtain the likelihood equations for $\pi_s (0)$, $\pi_s (A)$ and $\tau$ from the log-likelihood,
\begin{equation}
  \label{eq:20}
  \begin{aligned}
    l &= l(\pi_s (0), \pi_s (A), \tau) \\
    &=
    \sum_{i=1}^{N} \log \bigg\{
    (1-\tau)
    \exp \left\{
      \sum_{s=1}^{D} \left[ m_{is} \log \pi_s(0) + (T_{si} - m_{is}) \log (1-\pi_s(0)) \right]
    \right\} \\
    &\qquad +
    \tau
    \exp\left\{
      \sum_{s=1}^{D} \left[ m_{is} \log \pi_s(A) + (T_{si} - m_{is}) \log (1-\pi_s(A)) \right]
    \right\}
    \bigg\} \\
    &=
    \sum_{i=1}^{N} \log
    \left\{
    (1-\tau) e^{l_i (0)} + \tau e^{l_i (A)}
    \right\} ,
  \end{aligned}
\end{equation}
by differentiation. Note that $l_i (0)$ and $l_i (A)$ are the log-likelihoods for the accuracies of subject $i$ given the stimulus intensity $s$ and that $\alpha_i$ is $0$ or $A$, respectively. Thus,
\begin{equation}
  \label{eq:21}
  \begin{aligned}
    \frac{\partial}{\partial \pi_s (0)} l
    &=
    \sum_{i=1}^{N} \frac{
      (1-\tau) e^{l_i(0)} \frac{\partial}{\partial \pi_s (0)} l_i(0)
    }{
      (1-\tau) e^{l_i (0)} + \tau e^{l_i (A)}
    } \\
    &=
    \sum_{i=1}^{N}
    \frac{
      1
    }{
      1 + \frac{\tau}{1-\tau} e^{l_i (A) - l_i(0)}
    }
    \left\{
      \frac{m_{is}}{\pi_s(0)} - \frac{T_{is} - m_{is}}{1 - \pi_s(0)}
    \right\} \\
    &=
    \frac{1}{\pi_s(0)(1-\pi_s(0))}
    \sum_{i=1}^{N}
    \frac{
      m_{is} - \pi_s(0) T_{is}
    }{
      1 + \frac{\tau}{1-\tau} e^{l_i (A) - l_i(0)}
    } ,
  \end{aligned}
\end{equation}
and, analogously,
\begin{equation}
  \label{eq:22}
    \frac{\partial}{\partial \pi_s (A)} l
    =
    \frac{1}{\pi_s(A)(1-\pi_s(A))}
    \sum_{i=1}^{N}
    \frac{
      m_{is} - \pi_s(A) T_{is}
    }{
      1 + \frac{1-\tau}{\tau} e^{l_i (0) - l_i(A)}
    } .
\end{equation}
We notice that as $\tau$ goes to zero or one, one of the equations (\ref{eq:21}) and (\ref{eq:22}) will go to zero while the other goes to the usual score function for a binomial proportion. Moreover,
\begin{equation}
  \label{eq:23}
  \frac{\partial}{\partial \tau} l =
  \sum_{i=1}^{N}
  \frac{
    e^{l_i(A)} - e^{l_i(0)}
  }{
    (1-\tau) e^{l_i(0)} + \tau e^{l_i(A)}
  }
  =
  \sum_{i=1}^{N}
  \frac{
    1 - e^{l_i(0) - l_i(A)}
  }{
    \tau + (1-\tau) e^{l_i(0) - l_i(A)}
  } .
\end{equation}

\chapter{Conditional independence from DAGs}
\label{cha:cond-indep-from}

In the present appendix we illustrate an algorithm to determine conditional independence in directed acyclic graphs (DAGs) and illustrate this by arguing for the statement $Y_t \: \vert \: \bm{Y}_{-t},\bm{S}_{-t}, S_t, \alpha \sim Y_t \: \vert \: S_t, \alpha$ in the DAG in Figure \ref{fig:scheme2b}. The algorithm is described for example in \cite[][Section 3.2.2]{lauritzen_graphical_1996}, and may be summarised as follows supposing that we wish to evaluate a statement of the form ``$\mathcal{A}$ is conditionally independent from $\mathcal{B}$ given $\mathcal{C}$''.
\begin{enumerate}
\item Take the ancestral graph for the nodes involved in the conditional independence statement (i.e. the graph with the nodes $(\mathcal{A}, \mathcal{B}, \mathcal{C})$ and their ancestors).
\item Moralise the graph, marrying the parents. Delete the direction of the arrows.
\item In the moralised graph, delete the nodes that are conditioned on and all connections to these nodes to obtain the final graph. Note that in the statement ``$\mathcal{A}$ is conditionally independent from $\mathcal{B}$ given $\mathcal{C}$'' the nodes to delete are those in $\mathcal{C}$.
\item Conditional independence between $\mathcal{A}$ and $\mathcal{B}$ then holds if $\mathcal{A}$ and $\mathcal{B}$ are not connected in the final graph.
\end{enumerate}

\begin{figure}[htb]
  \centering
  \begin{tikzpicture}[node distance=2.5cm,auto,>=latex']
    \node [int, left of = empt2] (s0) {$S_{1}$};
    \node [int, below of = s0, node distance=2cm] (y0) {$Y_{1}$};
    \node [int, right of = empt2] (s1) {$S_{t-1}$};
    \node [int, below of = s1, node distance=2cm] (y1) {$Y_{t-1}$};
    \node [int, right of = s1, node distance=2cm] (s2) {$S_t$};
    \node [int, below of = s2, node distance=2cm] (y2) {$Y_t$};
    \node [left of = s1, node distance=2cm, minimum size=2em] (empt2) {\Huge $\ldots$};
    \node [int, circle, below of = empty2, node distance=2cm] (alph) {$\alpha$};
    \node [below of = empt2, node distance=2cm, minimum size=2em] (empty2) {};
    \path[->] (s0) edge node {} (y0);
    \path[->] (s0) edge node {} (empt2);
    \path[->] (s1) edge node {} (y1);
    \path[->] (s1) edge node {} (s2);
    \path[->] (y0) edge node {} (empt2);
    \path[->] (y1) edge node {} (s2);
    \path[->] (s2) edge node {} (y2);
    \path[->] (empt2) edge node {} (s1);
    \path[->] (empty2) edge node {} (s1);
    \path[->] (alph) edge node {} (y0);
    \path[->] (alph) edge node {} (y1);
    \path[->] (alph) edge node {} (y2);
  \end{tikzpicture}
  \caption{Ancestral graph for $\left( Y_t, \bm{Y}_{-t}, S_t, \bm{S}_{-t}, \alpha \right)$ from Figure \ref{fig:scheme2a}.}
  \label{fig:ancestral-graph}
\end{figure}
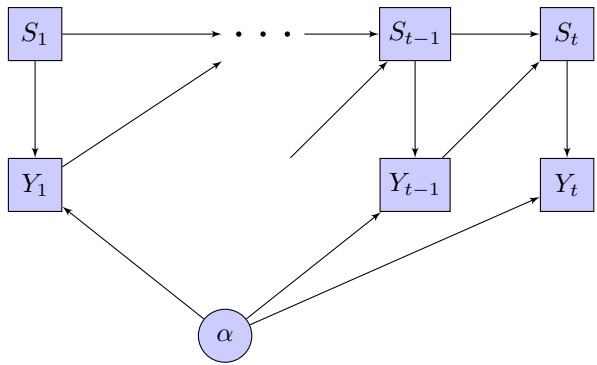

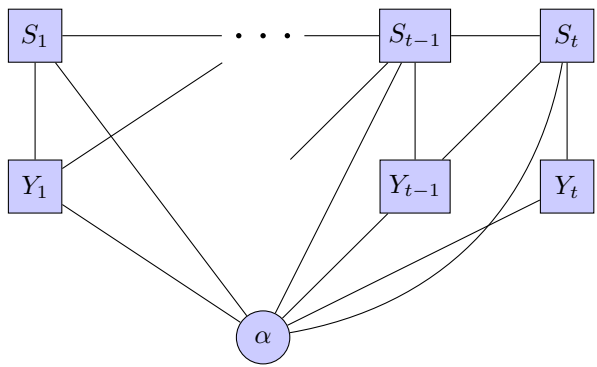
\begin{figure}[htb]
  \centering
  \begin{tikzpicture}[node distance=2.5cm,auto,>=latex']
    \node [int, left of = empt2] (s0) {$S_{1}$};
    \node [int, below of = s0, node distance=2cm] (y0) {$Y_{1}$};
    \node [int, right of = empt2] (s1) {$S_{t-1}$};
    \node [int, below of = s1, node distance=2cm] (y1) {$Y_{t-1}$};
    \node [int, right of = s1, node distance=2cm] (s2) {$S_t$};
    \node [int, below of = s2, node distance=2cm] (y2) {$Y_t$};
    \node [left of = s1, node distance=2cm, minimum size=2em] (empt2) {\Huge $\ldots$};
    \node [below of = empt2, node distance=2cm, minimum size=2em] (empty2) {};
    \node [int, circle, below of = empty2, node distance=2cm] (alph) {$\alpha$};
    \path[-] (s0) edge node {} (y0);
    \path[-] (s0) edge node {} (empt2);
    \path[-] (s1) edge node {} (y1);
    \path[-] (s1) edge node {} (s2);
    \path[-] (y0) edge node {} (empt2);
    \path[-] (y1) edge node {} (s2);
    \path[-] (s2) edge node {} (y2);
    \path[-] (empt2) edge node {} (s1);
    \path[-] (empty2) edge node {} (s1);
    \path[-] (alph) edge node {} (y0);
    \path[-] (alph) edge node {} (y1);
    \path[-] (alph) edge node {} (y2);
    \path[-] (alph) edge node {} (s0);
    \path[-] (alph) edge node {} (s1);
    \path[-, bend right = 35] (alph) edge node {} (s2);
  \end{tikzpicture}
  \caption{Moralised version of Figure \ref{fig:ancestral-graph}.}
  \label{fig:moral-graph}
\end{figure}

We illustrate the algorithm on the DAG from Figure \ref{fig:scheme2b} for the statement $Y_t \: \vert \: \bm{Y}_{-t},\bm{S}_{-t}, S_t, \alpha \sim Y_t \: \vert \: S_t, \alpha$, which is read ``$Y_t$ is conditionally independent from $\bm{Y}_{-t}$ and $\bm{S}_{-t}$ given $S_t$ and $\alpha$''. Figure \ref{fig:ancestral-graph} shows the ancestral graph for the nodes involved in the statement. A moralised version of the graph where the directions have been removed is given in Figure \ref{fig:moral-graph}.

To evaluate the statement concerning conditional independence we delete the nodes $S_t$ and $\alpha$ in Figure \ref{fig:moral-graph} and see that $Y_t$ is not connected to $\bm{Y}_{-t}$ and $\bm{S}_{-t}$ in the resulting graph, thus verifying the statement.

\end{document}